\documentclass[aps,prb,showpacs,twocolumn,floatfix,letterpaper,citeautoscript,10pt]{revtex4-1}

\usepackage[intlimits,sumlimits]{amsmath}
\usepackage{amsfonts,amssymb}
\usepackage{bm}
\usepackage{graphicx}
\usepackage{hyperref}
\usepackage[justification=centering]{caption}
\usepackage[vcentermath]{youngtab}
\usepackage{array}
\usepackage{multirow}

\newcommand{\be}{\begin{equation}}
\newcommand{\ee}{\end{equation}}
\providecommand{\e}[1]{\ensuremath{\times 10^{#1}}}

\newcommand*\pFqskip{8mu}
\catcode`,\active
\newcommand*\pFq{\begingroup
        \catcode`\,\active
        \def ,{\mskip\pFqskip\relax}%
        \dopFq
}
\catcode`\,12
\def\dopFq#1#2#3#4#5{%
        {}_{#1}F_{#2}\biggl[\genfrac..{0pt}{}{#3}{#4};#5\biggr]%
        \endgroup
}

\hypersetup{%
pdftitle={Landau Level Mixing and the Fractional Quantum Hall Effect},%
pdfauthor={Inti Sodemann, Allan H. MacDonald},%
pdfpagemode={UseNone},%
pdfstartview={FitH},%
breaklinks=true}

\begin{document}
\title{Landau Level Mixing and the Fractional Quantum Hall Effect}

\author{I. Sodemann}
\affiliation{Department of Physics, University of Texas at Austin, Austin, Texas 78712}
\author{A. H. MacDonald}
\affiliation{Department of Physics, University of Texas at Austin, Austin, Texas 78712}

\date{\today}

\begin{abstract}
We derive effective Hamiltonians for the fractional quantum Hall effect in 
$n=0$ and $n=1$ Landau levels that account perturbatively for Landau level mixing by 
electron-electron interactions.  To second order in the ratio of 
electron-electron interaction to cyclotron energy,
Landau level mixing is accounted for by constructing effective interaction 
Hamiltonians that include two-body and three-body contributions characterized by 
Haldane pseudopotentials.  
Our study builds upon previous treatments, using as a stepping stone the observation that
the effective Hamiltonian is fully determined by the few-body problem with  
$N=2$ and $N=3$ electrons in the partially filled Landau level.  
For the $n=0$ case we use a first quantization approach to 
provide a compact and transparent derivation of the effective Hamiltonian
which captures a class of virtual processes omitted in earlier 
derivations of Landau-level-mixing corrected Haldane pseudopotentials. 
\end{abstract}
\maketitle

\section{Introduction}
\label{sec:Introduction}

In a two-dimensional electron gas external magnetic fields form 
macroscopically degenerate groups of single-particle kinetic energy eigenstates 
known as Landau levels (LLs).  Many-electron state degeneracies grow 
exponentially with system size when these Landau levels are fractionally occupied and 
electron-electron interactions and disorder are ignored.  
In this limit the zero temperature chemical potential $\mu$ is constant over 
integer width intervals of the Landau level filling factor $\nu$,   
jumping between single-particle eigenenergies at 
integer values of $\nu$. 
($\nu \equiv N/N_{LL}$ where $N$ is the 
number of electrons in the gas and $N_{LL}$ is the number of single-particle states in each Landau level.)
The fractional quantum Hall effect~\cite{Tsui1982} is a consequence
of jumps in $\mu$ at non-integer values of $\nu$, and therefore can occur only in interacting electron systems.
  
Because of the exponential degeneracy of the many-body ground state when interactions are neglected, 
it is not possible to understand the fractional quantum Hall effect by treating interactions as 
a weak perturbation.  Instead, the problem of interactions in systems with partially 
filled Landau levels has traditionally been simplified by allowing occupation numbers to 
fluctuate only within the partially filled level.  This projection of the interaction Hamiltonian onto a single Landau 
has a long history in theories of two-dimensional electron systems, and
was first employed~\cite{Fukuyama1979,Yoshioka1979} even prior to the fractional quantum Hall
effect's discovery.\cite{Tsui1982} It is strictly justified as a low energy theory, only when the interaction energy per particle is small compared to the energetic separation between the Landau levels.  In this article we derive effective Hamiltonians which account for corrections to the projected Hamiltonian that are valid to leading order in interaction strength.
These corrections account for quantum fluctuations in otherwise empty and full Landau levels, and 
are therefore normally referred to as Landau level mixing corrections.     

Because of its non-perturbative character, the problem of 
interactions in a system with a partially filled Landau level has been a rich source of 
unique correlated electron phenomena, including fractional and non-Abelian quasiparticle
statistics~\cite{Nayak2008}, and electron-hole pair superfluidity~\cite{Eisenstein2004}. We limit our attention in this paper 
to the case of a two dimensional electron system
with parabolic bands.  The semiconductor quantum well systems in which the 
fractional quantum Hall effect has most often been observed are well 
described by such a model.  The small parameter on which our analysis is based is
the ratio of characteristic interaction and kinetic energy parameters in the strong
magnetic field limit of a two-dimensional
parabolic band:  

\be
\kappa\equiv \frac{e^2}{\hbar \omega_c \epsilon l_0},
\ee

\noindent where $\omega_c=eB/m^{*}c$ is the cyclotron frequency, $l_0=\sqrt{\hbar c/eB}$
is the magnetic length, $m^*$ is the parabolic band effective mass,
and $\epsilon$ is the low-frequency 
dielectric constant of the environment hosting the two dimensional electron system. (Hereafter $\hbar=1$.)
Note that $\kappa$ varies as $1/\sqrt{B}$, reaching its smallest values at the largest fields.  
The fact that the projected Hamiltonian is able to provide an adequate description of 
most properties of systems with fractionally filled Landau levels is, at first sight, somewhat surprising. 
In electron-doped gallium arsenide (GaAs), for example,
$\kappa \sim 0.4$ even at the highest achievable steady magnetic fields, $B\sim40 T$.
In hole-doped GaAs~\cite{Kumar2011}, AlAs,~\cite{Padmanabhan2010} and in the 
recently studied ZnO heterostructures~\cite{Maryenko2012} effective masses
are larger, reducing the cyclotron energy and increasing $\kappa$ further, as summarized in Table~\ref{tabkappa}. Landau level mixing is also expected to be substantial in graphene~\cite{Peterson2013} and in silicon quantum wells~\cite{Kott2012}.  Additionally, some of the most interesting fractional quantum Hall states occur in higher levels, and therefore at weaker magnetic fields and hence larger $\kappa$ values.  Our goals in this paper are to shed light on why the influence of LL mixing on the fractional quantum Hall effect is often modest, and to make progress in understanding its role when it is essential.

\begin{table}[b]
\caption{Relative strength of Coulomb and cyclotron
energies in GaAs, AlAs~\protect\cite{Padmanabhan2010} and ZnO~\protect\cite{Maryenko2012} ($B$ is measured in Tesla).} 
\centering 
\begin{tabular}{c  c  c  c} 
\hline\hline 
 & $m^*/m_0$ & $\epsilon$ & $\kappa$ \\ [0.5ex] 
\hline 
electron GaAs & \ $0.069$ \ & \ $13$ \ & $2.6/\sqrt{B}$ \\ 
hole GaAs & \ $0.39$ \ & \ $13$ \ & $14.6/\sqrt{B}$ \\
electron ZnO & \ $0.29$ \ & \ $8.5$ \ &  $16.7/\sqrt{B}$ \\
electron AlAs & \ $0.46$ \ & \ $10$ \ &  $22.5/\sqrt{B}$ \\ [1ex] 
\hline 
\end{tabular}
\label{tabkappa} 
\end{table}

There are many specific motivations for the study of interaction induced LL mixing.  
One is to attempt to bring theory and experiment into closer quantitative agreement in cases where 
the qualitative picture is already understood.  It is widely recognized, for example, 
that theoretically predicted FQHE gaps are invariably larger than the experimentally measured ones~\cite{Willett1988}.  
Although part of the discrepancy can be attributed to disorder and to finite quantum well widths~\cite{Morf2003,Wan2005}, LL mixing is
also expected to play an important role~\cite{Melik-Alaverdian1995,Yoshioka1984,YOSHIOKA1986,Yoshioka1986a,Sondhi1993,Murthy2002,Morf2003}. 

Another motivation is to study the emergence of phases which would otherwise be unstable.
Of particular interest is the potential role of LL mixing on the
stabilization of the Moore-Read Pfaffian state,~\cite{Moore1991} 
generally  
believed to describe the incompressible state observed at filling fraction $\nu=5/2$~\cite{Willett1987,Pan1999}. 
Indeed the Pfaffian and its particle-hole conjugate, the anti-Pfaffian, 
are distinct phases~\cite{Lee2007,Levin2007}, which are energetically degenerate in the absence of LL mixing.
Which of these two states appears experimentally is completely determined by the  
particle-hole symmetry breaking terms that Landau level mixing generates. 
This state selection property for half-filled LLs applies to any incompressible state that is not particle-hole 
invariant.  
Many numerical studies support the view that a state of the Moore-Read type is favored at $\nu=5/2$~\cite{Morf1998,Rezayi2000,Moller2008,Peterson2008b,Peterson2008,Peterson2008a,Wojs2009,Wang2009,Storni2010,Lu2010,Wojs2010},  
but the detailed form of the LL mixing would determine which of the two particle-hole mirror states is preferred. It is still unsettled which one of these two states is selected by the LL mixing induced by pure Coulomb interactions. One study has found the Pfaffian state to be favored~\cite{Wojs2010a}, while another has found the anti-Pfaffian to be favored~\cite{Rezayi2011}, although their detailed account for the LL mixing was different. Depending on the details of the confinement, the intersubband LL mixing could also play an important role~\cite{Papic2012}, specially in wide wells, where experimental studies have highlighted its influence of on the stability of the $\nu=5/2$ state~\cite{Liu2011}.  
 
Additionally, differences in gap sizes between filling factor $n+\nu$ and $n +1 -\nu$ 
(or $2n+\nu$ and $2n+2-\nu$ when spin is an active degree-of-freedom), observed conspicuously in GaAs~\cite{Eisenstein2009} and in large $\kappa$ systems like AlAs~\cite{Padmanabhan2010,Padmanabhan2010a} and hole-doped GaAs~\cite{Kumar2011}, require particle-hole symmetry breaking and hence 
LL mixing.  Finally, particle-hole asymmetries in critical densities for the emergence of charge density wave states
at low particle or hole densities~\cite{MacDonald1984} also reflect LL mixing.

In this article we will construct an effective Hamiltonian which accounts for
LL mixing to leading perturbative order in $\kappa$.  This many-body Hamiltonian when solved exactly will be able to predict energies, in units of $\omega_c$, correctly to order $\kappa^2$,
and the projected many-body wavefunctions into the Landau level of interest to order $\kappa$. 
It is interesting to note that the new energy scale appearing in our analysis, $\omega_c \kappa^2$, is independent of magnetic field strength (unlike the dominant interaction scale of the FQHE, $\omega_c \kappa$, which grows as $\sqrt{B}$), 
and it is twice the effective Rydberg of the parabolic band system~\cite{Sondhi1993}: 

\be
\omega_c \kappa^2 = \frac{m^* e^4}{\epsilon^2}.
\ee

One of the earliest studies to account numerically for LL mixing in the second Landau level was performed by Rezayi and Haldane~\cite{Rezayi1990}. Analytical perturbative studies of LL mixing due to interactions were first carried out in the lowest Landau level by Murthy and Shankar~\cite{Murthy2002}. These studies were subsequently extended to the second Landau level by Bishara and Nayak~\cite{Bishara2009}, employing an analysis similar to the renormalization group (RG) for fermions.  The present work complements these earlier papers
by presenting new methods of derivation, adding some new results,
and correcting some previous results. Our work has been developed essentially in parallel with two recent studies by Peterson and Nayak~\cite{Peterson2013}, and Simon and Rezayi~\cite{Simon2013}, and our findings are largely in agreement with these two studies. A key observation in our approach is that  
the effective many body Hamiltonian can be constructed by solving $N=2$ and 
$N=3$ few-body problems, which we use to simplify some derivations. We compute two-body~\cite{Haldane1983,Haldane1987} and three-body~\cite{Simon2007,Bishara2009,Davenport2012} generalized Haldane pseudopotentials for these interactions, which can be incorporated into many-body numerical diagonalization studies. Although the values we list here 
for these pseudopotentials are specialized to the case of 2D Coulomb interactions, we have derived analytic and semianalytic formulae for all the pseudpotentials valid for any rotationally invariant interaction.
Pseudopotential parameter values for more realistic interaction models which account for 
finite quantum well widths can be conveniently computed from these expressions.

Our paper is organized as follows. Section~\ref{secLLL} provides a compact
derivation of the effective Hamiltonian in the lowest Landau level ($n=0$ LL) 
that is based on a first quantization formalism,
and is valid for both bosons and fermions. 
In Sec.~\ref{secSLL} we construct the effective Hamiltonian for a partially filled first excited Landau level ($n=1$ LL),
and compute its two-body and three-body generalized Haldane pseudopotentials. 
In Sec.~\ref{secPolLLL} we derive an effective Hamiltonian valid for $ 1 < \nu < 2$
that is valid for the special case of maximally polarized electronic states 
in which the majority spin state is full and the minority spin state is partially occupied. 
We also compute the two and  three-body generalized Haldane pseudopotentials
appropriate for these effective Halmitonians.  In Sec.~\ref{secDisc} we summarize our findings and present conclusions. 
We have relegated discussions of some effective interaction properties and 
calculation details to a series of appendices.  

\section{Lowest Landau level}\label{secLLL}

\subsection{Many body Hamiltonian to order $\kappa^2$}\label{secLLLa}

We consider a two dimensional electron system subjected to a perpendicular magnetic field ${\bf B}=-B{\bf e_z}$. 
Measuring all energies in units of the cyclotron frequency and all lengths in units of the magnetic length, the single-particle
spectrum of a non-interacting disorder-free spinless system 
consists of discrete Landau levels with energies, $\varepsilon_n=n+1/2$~\cite{cond-mat/9410047,Giuliani2005}. 
The non-interacting Hamiltonian (including the Zeeman energy contribution) 
and the full Hamiltonian including interactions are given respectively by, 
\be\label{Hams}
\mathcal{H}_0=\sum_{i} \big( \hat{n}_i  +1/2 -  g \sigma^z_{i} \big), 
\ee
and
\be
\mathcal{H}=\mathcal{H}_0  + \kappa \sum_{i<j} v_{ij},
\ee

\noindent where $\hat{n}_i+1/2$ is the kinetic energy operator of particle $i$, $g=g_s m^*/2 m_e$ with $m_e$ the mass of the electron in vacuum and $g_s$ the effective g-factor of the host material, and $v_{ij}=1/|r_i-r_j|$ is the dimensionless Coulomb potential.

Unless the filling fraction $\nu$ is an integer, the non-interacting 
many-body eigenstates are degenerate. 
From degenerate state perturbation theory, the eigen-energies can be determined to order $\kappa$ by 
projecting the interaction term onto the degenerate manifold of non-interacting eigenstates with energy $E_0$,

\be
\mathcal{H}_1=E_0+\kappa  \sum_{i<j} \mathcal{P} v_{ij} \mathcal{P},
\ee

\noindent $\mathcal{H}_1$ is the Hamiltonian commonly employed to study the fractional quantum Hall effect. Employing conventional degenerate second order perturbation theory, the correction to the next order, $\kappa^2$  in energies and $\kappa$ in the projected wavefunctions, can be obtained from the effective Hamiltonian,

\be\label{H2}
\mathcal{H}_2=\mathcal{H}_1-\kappa^2  \sum_{\substack{
i<j\\
k<l}}\mathcal{P}v_{ij}\mathcal{P}_{\perp} \frac{1}{\mathcal{H}_0-E_0} \mathcal{P}_{\perp}v_{kl}\mathcal{P},
\ee
where $\mathcal{P}$ is the projector into the degenerate 
non-interacting ground state manifold of energy $E_0$, and 
$\mathcal{P}_\perp=1-\mathcal{P}$, is the projector onto its orthogonal complement. In the absence of Zeeman energy, $\mathcal{H}_2$ is the first quantization version of the Hamiltonian considered in Ref.~\onlinecite{Murthy2002}.

In the case of the $n=0$ LL, the degenerate manifold would be a subspace of the lowest kinetic energy eigenspace with a definite projection of the total spin along the z-axis, $S_z$, and Zeeman energy $E_0=-2g S_z$. Nevertheless, the energy denominator in Eq.~\eqref{H2} only includes the kinetic energy difference between the virtual excited states and the states in the degenerate manifold. This is a consequence of the conservation of $S_z$, because the virtual excitations produced by the interactions $v_{ij}$ and $v_{kl}$ do not change it, and therefore, the Zeeman energy disappears from this energy denominator. Consequently, for any $S_z$, we can write the second order correction to the effective Hamiltonian when the lowest-energy $n=0$ LL is partially filled, as,

\be\label{H2b}
\mathcal{H}_2=\mathcal{H}_1-\kappa^2  \sum_{\substack{
i<j\\
k<l}}\mathcal{P}v_{ij}\mathcal{P}_{\perp} \frac{1}{\hat{n}} \mathcal{P}_{\perp}v_{kl}\mathcal{P},
\ee 

\noindent with $\hat{n}=\sum_i \hat{n}_i$. 
There are three possibilities for the two pairs of indices $(i,j)$ and $(k,l)$ appearing in the sum in Eq.~\eqref{H2b};
they can share both particle indices ({\it i.e.} $i=k$ and $j=l$), they can share only one 
particle index while the other two are distinct, or they can share no indices. 
The last possibility does not contribute to $\mathcal{H}_2$ because the projection operators force
each virtually excited particle to decay back into the lowest LL after both 
interactions act.  The other two possibilities 
are non-vanishing and yield respectively two- and three-body effective interactions~\cite{Murthy2002,Bishara2009}.
Below we address the two-body effective interactions first.  

\subsection{Two-body interactions}\label{2bodyLLL}

The effective interactions implied by Eq.~\eqref{H2b} are independent of the state of the many body system in the $n=0$ LL.
We can therefore, without loss of generality, determine the effective Hamiltonian by considering only the 
few body $N=2$ and $N=3$ cases. 
The two body interactions can be written as,

\be\label{V2b}
\mathcal{V}^{2b}=\kappa\mathcal{P}v_{12}\mathcal{P}-\kappa^2  \mathcal{P}v_{12}\mathcal{P}_{\perp} \frac{1}{\hat{n}} \mathcal{P}_{\perp}v_{12}\mathcal{P}.
\ee

\noindent We have explicitly verified that this interaction is identical to the 
the two body interaction in Ref.~\onlinecite{Murthy2002}.  In particular, after translating
$\mathcal{V}^{2b}$ from Eq. ~\eqref{V2b} into its equivalent second quantized version, the piece for which both particles are virtually excited into higher Landau levels corresponds to the interaction $\delta H^2_{00}$ in 
Eq.~(26) of Ref.~\onlinecite{Murthy2002}, and the piece of Eq.~\eqref{V2b} for which only one particle is
excited into a higher Landau level corresponds, after normal ordering is performed, to the two body part obtained
from $\delta H^1_{00}$ in Eq.~(30) of Ref.~\onlinecite{Murthy2002}. 

The translational and rotational invariance of Eq.~\eqref{V2b} 
permits its decomposition into Haldane pseudopotentials. 
The virtual excitations are most 
easily analyzed by decomposing the degrees of freedom into center of mass, $R$, and relative, $r$, coordinates.
The total kinetic energy is then the sum of relative and center of mass kinetic energies, $\hat{n}=\hat{n}_1+\hat{n}_2=\hat{n}_r+\hat{n}_R$.  Because the interaction acts only on the relative 
coordinate, only $\hat{n}_r$ enters the energy denominator. 
We write the two-body Haldane pseudopotentials as $V^{2b}_m=\kappa V_1(m)+\kappa^2 V_2(m)$ where $V_1(m)$ 
are the standard Haldane pseudopotentials~\cite{Haldane1983,Haldane1987,cond-mat/9410047,Giuliani2005} given by 

\be\label{V1m}
V_1(m)={}_r\langle 0,m| v_{12} |0,m\rangle_r = \frac{\Gamma(m+1/2)}{2 m!}.
\ee

\noindent Here $|n,m\rangle_r$ is an state with the relative guiding center and kinetic energy quantum numbers 
$m$ and $n$ respectively. Note that the form of the relative motion eigenstates differ from the familiar single-particle 
eigenstates only because of the difference between relative 
motion and single-particle motion magnetic lengths $\ell_{r}= \sqrt{2} \ell_{0}$.  
For $V_2(m)$ we obtain:

\begin{multline}\label{V2m0}
V_2(m)=-\sum_{n=1}^{\infty}\frac{|{}_r\langle n,m+n| v_{12} |0,m\rangle_r|^2}{n}=\\
-\frac{[V_1(m)]^2}{4 (m+1)} \ \pFq{4}{3}{1,1,\frac{3}{2},\frac{3}{2}}{2,2,m+2}{1},
\end{multline}

\noindent with ${}_4F_3$ the generalized hypergeometric function.
Values of these pseudopotentials are presented in Table~\ref{tabVm2}, together with the usual Haldane pseudopotentials for comparison. These pseudopotentials agree with those computed in Refs.~\onlinecite{Peterson2013,Simon2013}.
Additionally very closely related numbers have been computed previously
in a strong magnetic-field expansion of the spectrum of hydrogenic atoms.\cite{MacDonald1986}
$V_2(m)$ equals $\pi \alpha_{0,m}/4$ where $\alpha_{0,m}$ values are specified by  
Eq.~(15) and Table II of Ref~\onlinecite{MacDonald1986}. 

\begin{table*}
\caption{Conventional two-body Haldane pseudopotentials in the $n=0$ LL ($V_1$ from Eq.~\eqref{V1m}), and the coefficients of their leading pertubative corrections ($V_2$ from Eq.\eqref{V2m0}).}  
\begin{tabular}{c  c  c  c  c  c  c  c  c  c  c} 
\hline\hline 
 $m$ & \ $0$ & $1$ & $2$ & $3$ & $4$ & $5$ & $6$ & $7$ & $8$ & $9$ \\[0.5ex] 
\hline
$V_1$ & \ $0.8862$ & $0.4431$ & $0.3323$ & $0.2769$ & $0.2423$ & $0.2181$ & $0.1999$ & $0.1856$ & $0.1740$ & $0.1644$ \\ [0.5ex] 
$V_2$ & \ $-0.3457$ & $-0.0328$ & $-0.0112$ & $-0.0055$ & $-0.0033$ & $-0.0022$ & $-0.0015$ & $-0.0012$ & $-0.0009$ & $-0.0007$ \\ [0.5ex] 
\hline\hline 
\end{tabular}
\label{tabVm2} 
\end{table*}

Our results for the leading order Haldane pseudopotential corrections differ from those presented in Table I of Ref.~\onlinecite{Bishara2009}.
The reason for this discrepancy is that the set of virtual processes in which only one particle is excited into higher Landau levels was omitted in Ref.~\onlinecite{Bishara2009}. This discrepancy has been recently solved in Ref.~\onlinecite{Peterson2013}. It was shown there that, in the RG language, these additional processes arise from keeping track of the correct normal ordering of the three-body interactions which give rise to additional contributions to the two-body interactions. 

\begin{table*}
\caption{Coefficients of the leading $S=3/2$ ($V_{3/2}$ from Eq.~\eqref{V3b3/2}) and $S=1/2$ ($V_{1/2}$ from Eq.~\eqref{V3b1/2}) three-body Haldane pseudopotentials in lowest LL. For $S=3/2$ there are two states with total angular momentum $m=9$ so the pseudopotential is a 
$2 \times 2$ matrix. For $S=1/2$ there are two states for $4\leq m\leq6$  and Haldane pseudopotentials are matrices, in these cases the listed 
pairs have been orthonormalized by rotating only the state with $\sigma=2$, as described in Appendix~\ref{3bapp}.} 
\centering 
\begin{tabular}{c  c  c  c  c  c  c} 
\hline\hline 
$m$ & \ $3$ & $5$ & $6$ & $7$ & $8$ & $9$ \\ [0.5ex] 
\hline 
 $(k,l)$ & \ $(0,1)$ & $(1,1)$ & $(0,2)$ & $(2,1)$ & $(1,2)$ & $(0,3) \  \ (3,1)$ \\ [0.5ex]
$V_{3/2}(k'l',kl)$ & \ $-0.0181$ & $0.0033$ & $-0.0107$ & $0.0059$ & $-0.0048$ & 
$\bigl(\begin{smallmatrix}
-0.0049& -0.0007\\
-0.0007& 0.0052
\end{smallmatrix}\bigr)$\\ [0.5ex] 
\hline\hline
 $m$ & \ $1$ & $2$ & $3$ & $4$ & $5$ & $6$ \\ [0.5ex] 
\hline 
$(\sigma,k,l)$ & \ $(1,0,0)$ & $(2,0,0)$ & $(1,1,0)$ & $(1,0,1) \ (2,1,0)$ & $(1,2,0) \   (2,0,1)$ & $(1,1,1) \ (2,2,0)$ \\ [0.5ex]
$V_{1/2}(\sigma'k'l',\sigma kl)$ & \ $-0.0345$ & $-0.0540$ & $0.0425$ & $\bigl(\begin{smallmatrix}
-0.0343& -0.0025\\
-0.0025& 0.0075
\end{smallmatrix}\bigr)$& 
$\bigl(\begin{smallmatrix}
0.0277& 0.0067\\
0.0067& -0.0176
\end{smallmatrix}\bigr)$& 
$\bigl(\begin{smallmatrix}
-0.0119& -0.0050\\
-0.0050& 0.0102
\end{smallmatrix}\bigr)$ \\  [1ex] 
\hline\hline 
\end{tabular}
\label{tabV3bLLL} 
\end{table*}

Since we have made no explicit reference to the statistics of the particles involved, our results apply equally well to fermions and bosons.  The difference between the two cases is only in the constraint imposed by quantum statistics on the allowed states, which implies that odd (even) $m$ pseudopotentials are associated with
spin triplet wavefunction for fermions (bosons) and spin 
singlets for bosons (fermions).  The case of a partially filled $n > 0$ LL, that we will discuss later in Sec.~\ref{secSLL}, is relevant only for fermions since partially filled higher Landau level states are not non-interacting ground states in the bosonic case.  

\subsection{Three-body interactions}
In order to derive the full Hamiltonian including the three body terms, it is 
sufficient to consider a $N=3$ few particle problem.  Three-body interactions follow from the 
terms in Eq.~\eqref{H2b} where only one index is shared between the pairs $(i,j)$ and $(k,l)$ while the other two 
indices are distinct, {\it e.g.} $i=k$ but $j\neq l$.  This contribution can be written as

\be\label{V3b}
\mathcal{V}^{3b}=-\kappa^2 \sum_{s\in S_3}\Pi_s \mathcal{P}v_{13}\mathcal{P}_{\perp} \frac{1}{\hat{n}} \mathcal{P}_{\perp}v_{12}\mathcal{P} \Pi_s,
\ee

\noindent where the sum is over the six permutations of three objects 
and $\Pi_s$ is the associtated unitary permutation operator. 
Since bosonic and fermionic states are both eigenstates of the permutation operator 
$\Pi_s|\Psi\rangle=\pm|\Psi\rangle$, it follows that 

\be\label{spin3pseudo}
\langle \Phi| \mathcal{V}^{3b}|\Psi\rangle=-6 \kappa^2 \langle \Phi| \mathcal{P}v_{13}\mathcal{P}_{\perp} \frac{1}{\hat{n}} \mathcal{P}_{\perp}v_{12}\mathcal{P}|\Psi\rangle, 
\ee

\noindent for arbitrary states $\{|\Psi\rangle,|\Phi\rangle\}$ with the same parity under permutations. 
Generalized Haldane pseudopotentials for $N$-body interactions have been thoroughly 
discussed for spinless particles by Simon, Rezayi, and Cooper in   Ref.~\onlinecite{Simon2007}, and for spinful particles by Davenport and Simon in Ref.~\onlinecite{Davenport2012}.  
We specialize hereafter in the case of fermions. There are two possible values for 
the total spin of three particles, namely $S=3/2$ and $S=1/2$.  The spatial wavefunction for 
$S=3/2$ must be fully antisymmetric, while the one for $S=1/2$ has mixed symmetry. 
For $S=3/2$ we employ the fully antisymmetric wavefunctions for three particles in the  lowest LL constructed by Laughlin in Ref.~\onlinecite{Laughlin1983}, whose polynomial part is

\be\label{Psi3/2}
\Psi^{\scriptscriptstyle{3/2}}_{kl}=\frac{1}{Z^{\scriptscriptstyle{3/2}}_{kl}}(z_a^2+z_b^2)^{k}\left[\frac{(z_a+iz_b)^{3 l}-(z_a-iz_b)^{3 l}}{2i}\right],
\ee

\noindent with $Z^{\scriptscriptstyle{3/2}}_{kl}=2^{3l+2k+1}[\pi^3(3l+k)! k!]^{1/2}$, $k\geq0$, $l\geq1$, and

\be
z_a=\sqrt{\frac{2}{3}}\left(\frac{z_1+z_2}{2}-z_3\right), \ \ z_b=\frac{z_1-z_2}{\sqrt{2}}.
\ee 

These wavefunctions form a complete orthonormal basis for the relative internal states of three particles, $\langle \Psi^{\scriptscriptstyle{3/2}}_{k'l'}|\Psi^{\scriptscriptstyle{3/2}}_{kl}\rangle=\delta_{k',k}\delta_{l',l}$.
The relative orbital angular momentum of the states is  $\hat{m}|\Psi^{\scriptscriptstyle{3/2}}_{kl}\rangle=(2 k+3 l)|\Psi^{\scriptscriptstyle{3/2}}_{kl}\rangle$. Therefore, it follows from the 
rotational invariance of interactions that the generalized Haldane pseudopotential matrix satisfies,

\be\label{V3b3/2}
\langle\Psi^{\scriptscriptstyle{3/2}}_{k'l'}|\mathcal{V}^{3b}|\Psi^{\scriptscriptstyle{3/2}}_{kl}\rangle=\kappa^2 \delta_{2k'+3l',2k+3l}V_{3/2}(k'l',kl).
\ee

In this way we find the $S=3/2$ pseudopotentials listed in Table~\ref{tabV3bLLL}. These pseudopotentials agree with those derived in Refs.~\onlinecite{Bishara2009,Peterson2013}. They are also in close agreement with the pseudopotential differences obtained in Ref.~\onlinecite{Rezayi2011}.
Further details on the properties of these states and the derivation of these 
pseudopotential values can be found in Appendix~\ref{3bapp}.

To construct the $S=1/2$ states we follow the approach of Davenport and Simon~\cite{Davenport2012}. We start with two primitive polynomials which are antisymmetric only under permutations of variables $1$ and $2$, and thus have the symmetry of the Young tableau~\cite{Hamermesh1989}

\be\label{yng}
\young(13,2) \ .
\ee

\noindent These primitive polynomials are,

\be
\beta_1=z_b, \ \beta_2=z_az_b.
\ee

\noindent The most general polynomials with the symmetry of this tableau are then obtained by multiplying the primitive polynomials by the most general fully symmetric translationally invariant polynomial. We choose a different basis for the fully symmetric polynomials from that employed in Ref.~\onlinecite{Davenport2012}, which makes calculations simpler. Our basis for the symmetric polynomilas is the bosonic analog of the fermionic wavefunctions of Ref.~\onlinecite{Laughlin1983}. By multiplying this bosonic wavefunction by the primitive polynomials, $\beta_\sigma$, we obtain the polynomial part of the $S=1/2$ fermionic wavefunctions,

\be\label{S=3/2Psi}
\Psi^{\scriptscriptstyle{1/2}}_{\sigma kl}=\frac{\beta_\sigma}{Z^{\scriptscriptstyle{1/2}}_{\sigma kl}}(z_a^2+z_b^2)^{k}\left[\frac{(z_a+iz_b)^{3 l}+(z_a-iz_b)^{3 l}}{2}\right],
\ee

\noindent with $\sigma=\{1,2\}$, $k\geq0$, $l\geq0$. The normalization factor for $\sigma=1$ states is given by $Z^{\scriptscriptstyle{1/2}}_{1kl}=Z^{\scriptscriptstyle{3/2}}_{kl}  \sqrt{(1+ \delta_{l,0})(2k+2+3l)}$, and for $\sigma=2$ it is  $Z^{\scriptscriptstyle{1/2}}_{2kl}=Z^{\scriptscriptstyle{3/2}}_{kl}
\sqrt{(1+ \delta_{l,0})[(k+1+3l)(k+2+3l)+(k+1)(k+2)]}$, where $Z^{\scriptscriptstyle{3/2}}_{kl}$ is the normalization constant of the $\Psi^{\scriptscriptstyle{3/2}}_{kl}$ states appearing in Eq.~\eqref{Psi3/2}.

The spin part of the $S=1/2$ wavefunction has the symmetry of the Young tableau conjugate to~\eqref{yng}, thus the fully antisymmetric wavefunction is~\cite{Davenport2012},

\be
|\Psi^{\scriptscriptstyle{1/2}}_{\sigma kl}\rangle=\mathcal{A}\{\Psi^{\scriptscriptstyle{1/2}}_{\sigma kl}\otimes |\uparrow\uparrow\downarrow\rangle\},
\ee

\noindent where $\mathcal{A}=(1-\Pi_{23}+\Pi_{12}\Pi_{23})/\sqrt{3}$, is the partial 
antisymmetrization operator between $\uparrow$ and $\downarrow$ particles, 
and $\Pi_{ij}$ is the unitary operator corresponding to elementary permutations between $i$ and $j$. 
The states $|\Psi^{\scriptscriptstyle{1/2}}_{\sigma kl}\rangle$ are linearly indepenent 
and complete but not orthogonal:

\begin{widetext}
\be\label{innerproducts}
\begin{split}
& \langle \Psi^{\scriptscriptstyle{1/2}}_{\sigma k'l'}|\Psi^{\scriptscriptstyle{1/2}}_{\sigma kl}\rangle =\delta_{k',k}\delta_{l',l} \\
\langle \Psi^{\scriptscriptstyle{1/2}}_{1 k'l'}|\Psi^{\scriptscriptstyle{1/2}}_{2 kl}\rangle =-\pi^3 \frac{2^{4k'+6l'+1}}{Z^{\scriptscriptstyle{1/2}}_{1k'l'}Z^{\scriptscriptstyle{1/2}}_{2kl}} [k!  (k'+& 3 l')! \delta_{l',l+1}\delta_{k,k'+1}(1+\delta_{l,0})+k'! (k+3 l)! \delta_{l,l'+1}\delta_{k',k+2}(1+\delta_{l',0})].
\end{split}
\ee
\end{widetext}

\noindent The internal motion orbital angular momentum of these 
states is $\hat{m}|\Psi^{\scriptscriptstyle{1/2}}_{\sigma kl}\rangle=(\sigma+2 k+3 l)|\Psi^{\scriptscriptstyle{1/2}}_{\sigma kl}\rangle$. Thus the generalized Haldane pseudopotentials satisfy

\be\label{V3b1/2}
\langle\Psi^{\scriptscriptstyle{1/2}}_{\sigma' k'l'}|\mathcal{V}^{3b}|\Psi^{\scriptscriptstyle{1/2}}_{\sigma kl}\rangle=\kappa^2 \delta_{\sigma'+2k'+3l',\sigma+2k+3l}V_{1/2}(\sigma'k'l',\sigma kl),
\ee

\noindent and have the leading values listed in Table~\ref{tabV3bLLL}. These pseudopotentials are 
also in agreement with those derived in Refs.~\onlinecite{Bishara2009,Peterson2013} where the two leading three-body pseudopotentials for $S=1/2$ were computed.
Further details on these states and the $S=1/2$ three-body pseudopotentials can be found in Appendix~\ref{3bapp}.

\section{$n=1$ Landau level}\label{secSLL}

\subsection{Many body Hamiltonian to order $\kappa^2$}

The first quantization analysis presented in section~\ref{secLLL} for a partially filled lowest LL is
cumbersome in the case of a partially filled $n=1$ 
Landau level because of the need to account for virtual excitations of the full $n=0$ LL. 
In this section we therefore use a second quantization
approach.  By following a path entirely analogous to that of section~\ref{secLLLa}, 
we arrive at an effective Hamiltonian to order $\kappa^2$ which acts in the $n = 1$ LL.

The first order term is as usual simply the projection of the Hamiltonian onto the partially filled level,
\be
\mathcal{H}_1=E_0+\frac{\kappa}{2}  \sum_{\mu\nu m} v_{1 2 , 3 4} \ c^\dagger_{1 m_1 \nu} c^\dagger_{1 m_2 \mu} c_{1 m_3 \mu} c_{1 m_4 \nu},
\ee
where $c^\dagger_{n m \mu}$ is a fermion creation operator in LL $n$ with guiding center quantum number $m$, $\mu$ runs over all the fermion flavors, and $E_0$ is the non-interacting energy including the Zeeman term.  For generality we assume $f$-flavors ({\it i.e.} for spin $s$, $f=2s+1$). To order $\kappa^2$ the effective Hamiltonian is 

\begin{multline}\label{H2SLL}
\mathcal{H}_2=\mathcal{H}_1-\frac{\kappa^2}{4}  \sum_{\substack{
1\cdots8 \\
\mu\nu\lambda\sigma}} v_{1 2 , 3 4}  v_{5 6 , 7 8}\\
\times P_0 c^\dagger_{5 \nu} c^\dagger_{6 \mu} c_{7 \mu} c_{8 \nu} P_\perp \frac{1}{\hat{n}-N_1} P_\perp c^\dagger_{1 \lambda} c^\dagger_{2 \sigma} c_{3 \sigma} c_{4 \lambda} P_0,
\end{multline}

\noindent where the integers abbreviate single-particle kinetic and guiding center quantum numbers ({\it e.g.} $\{1 \Leftrightarrow n_1m_1 \}$),  $N_1$ is the number of particles in the 
partially filled $n=1$ LL, $P_0$ is the projector into the many-body eigenspace with $\hat{n}=N_1$, and $P_\perp$ is 
the projector into its orthogonal complement.~\footnote{As in the $n=0$ LL case, the derivation is independent of the presence of the Zeeman energy term, because virtual transitions into states with different $S_z$ are forbidden. This conclusion still holds for $f$ flavors and SU($f$)-invariant interactions in the presence of single particle Zeeman-like terms.}

The classification of all the interaction terms arising from this Hamiltonian is a lengthy bookeeping exercise that we describe in Appendix~\ref{intSLL}. There are no four body terms, and the one body terms, which we do not compute, account only for exchange interactions with the full Landau level.  These produce only a well known overall constant shift of the single-particle energies, that is equivalent to a change in the chemical potential~\cite{Rezayi1990}. 

\begin{figure}[t]
	\begin{center}
		\includegraphics[width=3.3in]{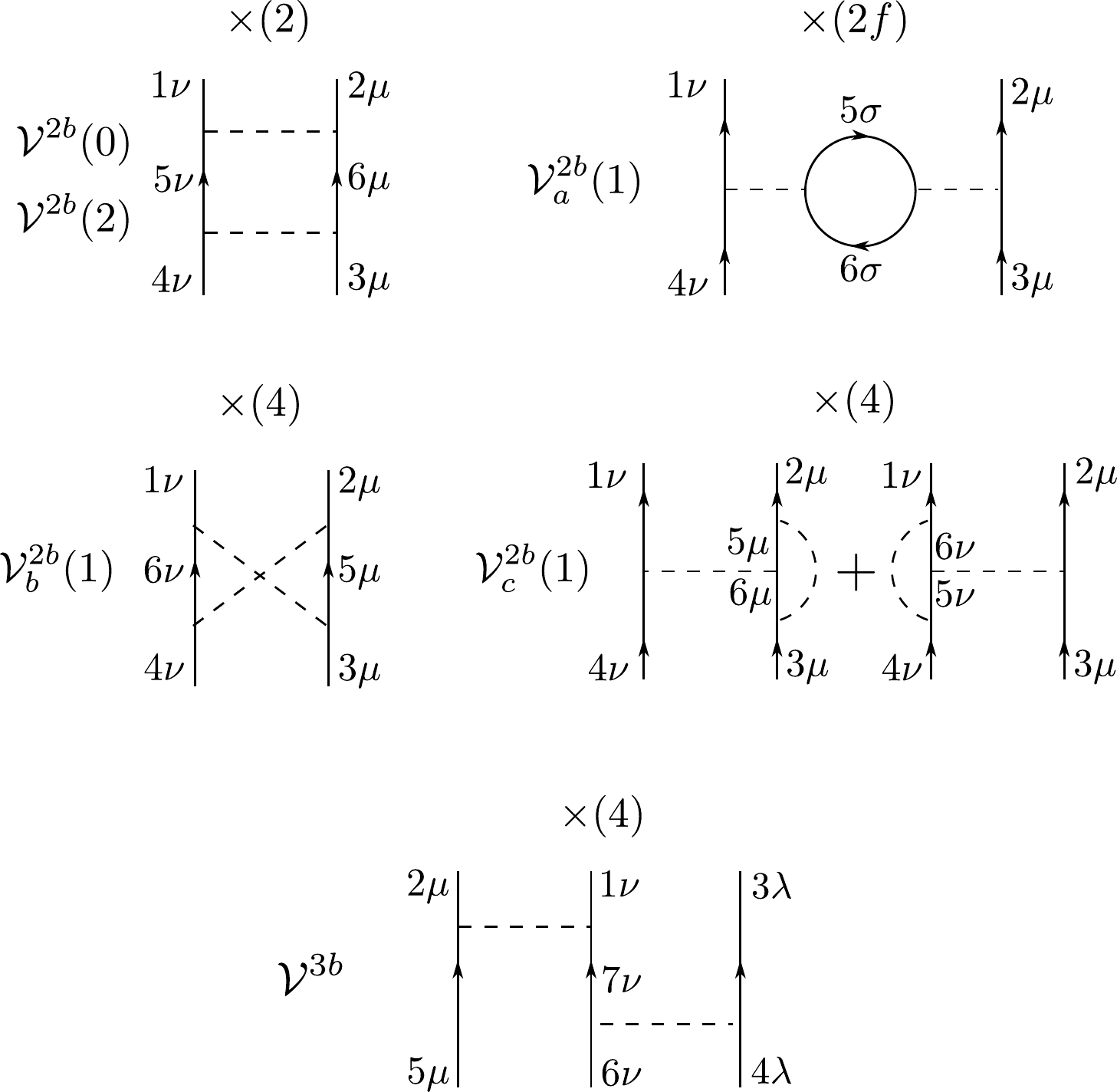}
	\end{center}
	\caption{Diagrams representing two- and three-body interactions at order $\kappa^2$ in the $n=1$ LL. 
	The symbol $\times (k)$ appearing above each diagram represents its multiplicity, {\it i.e.} there are $k$ different contractions of the operators in Eq.~\eqref{H2SLL} which give rise to the same diagram after relabeling dummy indices. The indices of the fermion lines are consistent with those appearing in Eqs.~\eqref{V2bSLL},~\eqref{V2bSLL2} and~\eqref{V3bSLL}.}
	\label{figdiagrams}
\end{figure}

The non-vanishing two-body terms satisfy a kinetic energy balance condition, $n_5+n_6=n_3+n_4$, and can take values $n_3+n_4=\{2,1,0\}$ in Eq.~\eqref{H2SLL}. We employ the three possible values of this 
{\it incoming} kinetic energy as labels for the three allowed terms, labeled as $\{\mathcal{V}^{2b}(2),\mathcal{V}^{2b}(1),\mathcal{V}^{2b}(0)\}$,
and find that:

\be\label{V2bSLL}
\begin{split}
\mathcal{V}^{2b}(2) &=
-\frac{\kappa^2}{2}  \sum_{\substack{
1\cdots6 \\ \mu\nu}
} v_{1 2 , 6 5}  v_{5 6 , 3 4} \frac{\theta(n_5,n_6)}{n_5+n_6-2} c^\dagger_{1 \nu} c^\dagger_{2 \mu} c_{3 \mu} c_{4 \nu}, \\
\mathcal{V}^{2b}(0) &=
-\frac{\kappa^2}{2}  \sum_{\substack{
1\cdots6 \\ \mu\nu}
} v_{1 2 , 6 5}  v_{5 6 , 3 4} \frac{\delta(n_5,n_6)}{2} c^\dagger_{1 \nu} c^\dagger_{2 \mu} c_{3 \mu} c_{4 \nu},
\end{split}
\ee

\noindent where the indices of the creation/annihilation operators are understood to be on the 
$n=1$ LL, $\theta(n_5,n_6)$ is a function that restricts $n_5+n_6\geq3$ and $n_5\geq1,n_6\geq1$, and $\delta(n_5,n_6)$ restricts $n_5=n_6=0$.  The term $\mathcal{V}^{2b}(1)$ can be split into three terms,

\be\label{V2bSLL2}
\begin{split}
\mathcal{V}^{2b}(1) &= \mathcal{V}^{2b}_a(1)+\mathcal{V}^{2b}_b(1)+\mathcal{V}^{2b}_c(1),\\
\mathcal{V}^{2b}_a(1) &=
- f \kappa^2 \sum_{\substack{
1\cdots6 \\ \mu\nu}
} v_{1 5 , 6 4}  v_{2 6 , 5 3} \frac{\tau(n_5,n_6)}{n_6} c^\dagger_{1 \nu} c^\dagger_{2 \mu} c_{3 \mu} c_{4 \nu},\\
\mathcal{V}^{2b}_b(1) &=
\kappa^2  \sum_{\substack{
1\cdots6 \\ \mu\nu}
} v_{1 5 , 3 6}  v_{6 2 , 5 4} \frac{\tau(n_5,n_6)}{n_6} c^\dagger_{1 \nu} c^\dagger_{2 \mu} c_{3 \mu} c_{4 \nu},\\
\mathcal{V}^{2b}_c(1) &=
\kappa^2  \sum_{\substack{
1\cdots6 \\ \mu\nu}
} (v_{1 5 , 4 6}  v_{6 2 , 3 5}+v_{5 1 , 4 6}  v_{2 6 , 3 5}) \\ & \quad \quad \quad \quad \times\frac{\tau(n_5,n_6)}{n_6} c^\dagger_{1 \nu} c^\dagger_{2 \mu} c_{3 \mu} c_{4 \nu},
\end{split}
\ee

\noindent where $\tau(n_5,n_6)$ restricts $n_5=0$ and $n_6\geq1$. Each of these two-body interactions has a simple diagrammatic representation.
In particular $\mathcal{V}^{2b}(2)$ and $\mathcal{V}^{2b}(0)$ correspond to particle-particle ladder diagrams,  
and $\mathcal{V}^{2b}_a(1)$ is a screening diagram, hence the flavor multiplicity factor.
$\mathcal{V}^{2b}_b(1)$ is a particle-hole ladder diagram, and $\mathcal{V}^{2b}_c(1)$ is
 a vertex correction diagram as summarized in Fig.~\ref{figdiagrams}. To get a feeling for the significant additional complication of the $n=1$ LL effective interaction calculation, note that from all of these terms only the particle-particle ladder type diagram is present in the $n=0$ LL.   When our first quantization formulation approach was used for the two body interactions
in the $n=0$ LL case, the problem was reduced to the simplicity of a one-body second-order perturbation theory calculation.  

By comparison, the three-body interactions are simpler.  They 
can be combined into a single term, 

\be\label{V3bSLL}
\mathcal{V}^{3b} =
-\kappa^2  \sum_{\substack{
1\cdots7 \\ \lambda\mu\nu}
} v_{1 2 , 5 7}  v_{7 3 , 4 6} \frac{\omega(n_7)}{n_7-1} c^\dagger_{1 \nu} c^\dagger_{2 \mu} c^\dagger_{3 \lambda} c_{4 \lambda} c_{5 \mu} c_{6 \nu},
\ee 

\noindent where again the Landau level indices of the 
operators are understood to have $n=1$ and $\omega(n_7)$ restricts $n_7\neq1$. 
The three-body interaction has a simple diagramatic representation depicted in Fig.~\ref{figdiagrams}. There is a compact first quantized version of Eq.~\eqref{V3bSLL}, which for three particles in the $n=1$ LL is

\be\label{V3bSLL2}
\mathcal{V}^{3b}=-\kappa^2 \sum_{s\in S_3}\Pi_s \mathcal{P}v_{13}\mathcal{P}_{\perp} \frac{1}{\hat{n}-3} \mathcal{P}_{\perp}v_{12}\mathcal{P} \Pi_s,
\ee

\noindent where the notation is the same as that of Eq.~\eqref{V3b}, except that $\mathcal{P}$ is the projector into the eigenspace with $\hat{n}=\hat{n}_1+\hat{n}_2+\hat{n}_3= 3$, 
and $\mathcal{P}_{\perp}$ into its orthogonal complement. This first quantized version significantly simplifies the evaluation of three-body Haldane pseudopotentials which we will discuss later on.

$\mathcal{V}^{2b}(2)$, $\mathcal{V}^{2b}_a(1)$, $\mathcal{V}^{2b}_b(1)$, $\mathcal{V}^{2b}_c(1)$, $\mathcal{V}^{2b}(0)$, and $\mathcal{V}^{3b}$, are separately rotationally and translationally invariant, as discussed in Appendix~\ref{2bpseud}.
This property guarantees that the interactions can be represented by Haldane pseudopotentials
as in the $n=0$ LL case.  
To compute the two-body Haldane pseudopotentials, we consider the $N=f N_{LL} +2$-body problem
in which the lowest Landau level is completely full with $f N_{LL} $ particles and only two particles are in the 
$n=1$ LL. 
The state describing the orbital part of the relative motion of two particles in the $n=1$ LL, $|m\rangle_1$, 
can be written as~\cite{Haldane1987},

\be\label{m1SLL}
|m\rangle_1=a^\dagger_1 a^\dagger_2 |0,m\rangle_r,
\ee

\noindent where $|m,0\rangle_r$ is the state of relative angular momentum $m$, of two particles in the 
$n=0$ LL discussed next to Eq.~\eqref{V1m}, and $a^\dagger_i$ is the kinetic energy raising operator for particle $i$.
Note that $|m\rangle_1$ is an entagled state of relative and center of mass quantum numbers as discussed in Appendix~\ref{2bpseud}. 
The usual Haldane pseudopotentials for the $n=1$ LL are~\cite{Haldane1987,Rezayi2000}

\begin{multline}\label{V1SLL}
V_1(m)={}_1\langle m|v_{12}|m\rangle_1=\\
\frac{\Gamma(m+1/2)}{2 m!}\frac{(m-3/8)(m-11/8)}{(m-1/2)(m-3/2)}.
\end{multline}

\noindent The Haldane pseudopotentials to order $\kappa^2$ are listed in Table~\ref{tabVm2SLL}. Our pseudopotentials are in agreement with those obtained in Refs.~\onlinecite{Rezayi1990,Peterson2013,Simon2013}, with small discrepancies pressumably arising from numerical error. We believe our pseudopotentials are essentially free from numerical errors because we have converted the effective interaction expressions to first quantization, as dicussed in Appendix~\ref{2bpseud}, which allows for very efficient calculations. We have explicitly verified that our two body interactions are equivalent to those of Ref.~\onlinecite{Peterson2013}. More specifically, the sum of our ladder type interactions from Eq.~\eqref{V2bSLL}, namely $\mathcal{V}^{2b}(2)+\mathcal{V}^{2b}(0)$, is the same as the sum of the BCS interaction of Eq. (19), with the two-body interaction appearing in the last line of Eq. (26) in Ref.~\onlinecite{Peterson2013}. And the sum of the interactions appearing in Eq.~\eqref{V2bSLL2}, namely $\mathcal{V}^{2b}_a(1)+\mathcal{V}^{2b}_b(1)+\mathcal{V}^{2b}_c(1)$, is identical to the sum of the ZS and ZS' interactions appearing in Eqs. (17) and (18), with the remainding two-body interactions appearing in the second and third line of Eq. (26) in Ref.~\onlinecite{Peterson2013}.  

It is interesting to note from Table~\ref{tabVm2SLL} that the leading two-body pseudopotentials in the $n=1$ LL, namely those with $m\leq3$, are dominated by the screening interaction $\mathcal{V}^{2b}_a(1)$. The contributions from the remaining interactions to these pseudopotentials nearly cancel. In Appendix~\ref{2bpseud} we discuss further properties of $\mathcal{V}^{2b}_a(1)$, and explicitly show that it is equivalent to the $\kappa^2$ term in the RPA approximation for the statically screened potential in the presence of a completelly filled $n=0$ LL.

\begin{table*}[h,t]
\caption{Conventional two-body Haldane pseudopotentials in 
the $n=1$ LL ($V_1$ from Eq.~\ref{V1SLL}) and coefficients of their leading perturbative corrections  for a completely filled spin-$1/2$ $n=0$ LL (from Eqs.~\eqref{V2bSLL} and~\eqref{V2bSLL2} with $f=2$), which describe the 
physics for filling factors $\nu$ in the interval $2 < \nu < 4$. The last row is the sum of all the perturbative corrections, {\it i.e.} $V^{2b}=V^{2b}(2)+V^{2b}_a(1)+V^{2b}_b(1)+V^{2b}_c(1)+V^{2b}(0)$.} 
\centering 
\begin{tabular}{c  c  c  c  c  c  c  c  c  c  c} 
\hline\hline 
 $m$ & \ $0$ & $1$ & $2$ & $3$ & $4$ & $5$ & $6$ & $7$ & $8$ & $9$ \\ [0.5ex] 
\hline 
$V_1$ & \ $0.6093$ & $0.4154$ & $0.4500$ & $0.3150$ & $0.2635$ & $0.2322$ & $0.2101$ & $0.1935$ & $0.1803$ & $0.1696$\\
$V^{2b}(2)$ & \ $-0.0903$ & $-0.0347$ & $-0.1235$ & $-0.0241$ & $-0.0110$ & $-0.0064$ & $-0.0042$ & $-0.0030$ & $-0.0022$ & $-0.0017$\\
$V^{2b}_a(1)$ & \ $-0.3930$ & $-0.2038$ & $-0.1981$ & $-0.1119$ & $-0.0535$ & $-0.0235$ & $-0.0098$ & $-0.0039$ & $-0.0015$ & $-0.0006$\\
$V^{2b}_b(1)$ & \ $0.0247$ & $0.0706$ & $0.0803$ & $0.0186$ & $-0.0031$ & $0.0091$ & $-0.0004$ & $0.0029$ & $0.0007$ & $0.0011$ \\
$V^{2b}_c(1)$ & \ $0.0750$ & $-0.0475$ & $0.0870$ & $0.0141$ & $-0.0128$ & $-0.0167$ & $-0.0139$ & $-0.0104$ & $-0.0077$ & $-0.0058$  \\
$V^{2b}(0)$  & \ $0$ & $0$ & $-0.0276$ & $-0.0023$ & $-0.0006$ & $-0.0003$ & $-0.0001$ & $-7\e{-5}$ & $-5\times10^{-5}$ & $-3\times10^{-5}$\\ [0.5ex] 
$V^{2b}$  & \ $-0.3836$ & $-0.2155$ & $-0.1818$ & $-0.1056$ & $-0.0810$ & $-0.0377$ & $-0.0285$ & $-0.0146$ & $-0.0108$ & $-0.0070$\\ [1ex] 
\hline\hline 
\end{tabular}
\label{tabVm2SLL} 
\end{table*}

The three-body states of the $N= f N_{LL} +3$ body problem
in the $n=1$ LL are mapped from those in the $n=0$ LL by raising the kinetic energy of the three particles,

\be
\begin{split}
|\Psi^{\scriptscriptstyle{3/2}}_{kl}\rangle_1&=a^\dagger_1a^\dagger_2a^\dagger_3|\Psi^{\scriptscriptstyle{3/2}}_{kl}\rangle, \\
|\Psi^{\scriptscriptstyle{1/2}}_{\sigma kl}\rangle_1&=a^\dagger_1a^\dagger_2a^\dagger_3|\Psi^{\scriptscriptstyle{1/2}}_{\sigma kl}\rangle.
\end{split}
\ee

\noindent With this construction the orthonormality of the $S=3/2$ states in $n=0$ LL and Eqs.~\eqref{V3b3/2},\eqref{innerproducts}, and \eqref{V3b1/2} are immediately extended to the $n=1$ LL three body states. 
The three body Haldane pseudopotentials we obtain are listed in Table~\ref{tabV1/2SLL}.
Further properties of these states and of the corresponding pseudopotentials are given in Appendix~\ref{3bapp}. The three-body Haldane pseudopotentials in the $n=1$ LL are in  agreement with those computed in Refs.~\onlinecite{Bishara2009,Peterson2013}. There is a discrepancy with those obtained numerically in Ref.~\onlinecite{Simon2013} which is likely to arise from the errors associated with the finite size effects and Landau level trunctaion present in such study.

\begin{table*}
\caption{Coefficients of the second order correction to the 
 $S=3/2$ and $S=1/2$ three-body Haldane pseudopotentials in the 
 $n=1$ LL. For $S=1/2$ and $4\leq m\leq6$ the pairs listed have been orthonormalized by rotating only the state with $\sigma=2$, as described in Appendix~\ref{3bapp}.} 
\centering 
\begin{tabular}{c  c  c  c  c  c  c} 
\hline\hline 
$m$ & \ $3$ & $5$ & $6$ & $7$ & $8$ & $9$ \\ [0.5ex] 
\hline 
 $(k,l)$ & \ $(0,1)$ & $(1,1)$ & $(0,2)$ & $(2,1)$ & $(1,2)$ & $(0,3) \ \ (3,1)$ \\ [0.5ex]
$V_{3/2}(k'l',kl)$ & \ $-0.0147$ & $-0.0054$ & $-0.0099$ & $0.0005$ & $-0.0009$ & 
$\bigl(\begin{smallmatrix}
-0.0088& 0.0007\\
0.0007& 0.0033
\end{smallmatrix}\bigr)$\\ [0.5ex] 
\hline\hline
 $m$ & \ $1$ & $2$ & $3$ & $4$ & $5$ & $6$ \\ [0.5ex] 
\hline 
$(\sigma,k,l)$ & \ $(1,0,0)$ & $(2,0,0)$ & $(1,1,0)$ & $(1,0,1) \ (2,1,0)$ & $(1,2,0) \ (2,0,1)$ & $(1,1,1) \ (2,2,0)$ \\ [0.5ex]
$V_{1/2}(\sigma'k'l',\sigma kl)$ & \ $-0.0319$ & $-0.0305$ & $-0.0131$ & 
$\big(\begin{smallmatrix}
-0.0009& -0.0004\\
-0.0004& -0.0100
\end{smallmatrix}\big)$ & 
$\big(\begin{smallmatrix}
-5\times10^{-5}& -0.0056\\
-0.0056& 0.0229
\end{smallmatrix}\big)$ & 
$\big(\begin{smallmatrix}
0.0067& 0.0017\\
0.0017& -0.0010
\end{smallmatrix}\big)$ \\  [1ex] 
\hline\hline 
\end{tabular}
\label{tabV1/2SLL} 
\end{table*}

\section{Partially filled spin polarized LLL}\label{secPolLLL}

Another instance for which a useful effective Hamiltonian can be obtained using the line of reasoning 
presented in the previous sections, is the case of a Landau level for which spin is a good quantum number, 
and $f_o$ spin states are completely filled ($f_o<f$) while the remaining $f-f_o$ spin states are partially empty. 
This approach is useful for electrons, for example, in addressing those states at filling factors in 
the interval $1 < \nu < 2$ for which all $\uparrow$ $n=0$ states are occupied, {\it i.e.} for 
maximally spin-polarized states.   Although we could use the effective Hamiltonian discussed in Sec.~\ref{secLLL} 
in this filling factor range, it is useful to derive an 
effective Hamiltonian which acts only on $\downarrow$ degrees of freedom, since the conservation of $S_z$ 
prevents the $\uparrow$ spins from participating in the low energy dynamics.  We emphasize that the 
ground state in the filling factor range $1 < \nu < 2$ is {\it not} always maximally spin-polarized, so this approach 
cannot always be used to describe the ground state.

\begin{table*}[ht]
\caption{Coefficients of the second order corrections to the two-body Haldane pseudopotentials for spin-$1/2$ fermions in a state with completely filled majority spins and partially empty minority spins in the $n=0$ LL.} 
\centering 
\begin{tabular}{c  c  c  c  c  c  c} 
\hline\hline 
 $m$ & \ $0$ & $1$ & $2$ & $3$ & $4$ & $5$ \\ [0.5ex] 
\hline 
$V^{2b}_o(m)$ & \ $-0.3662$ & $-0.0959$ & $-0.0268$ & $-0.0078$ & $-0.0023$ & $-0.0007$ \\ 
$V_2(m,1)$ & \ $-0.7119$ & $-0.1287$ & $-0.0380$ & $-0.0133$ & $-0.0056$ & $-0.0029$ \\ [1ex] 
\hline\hline 
\end{tabular}
\label{tabVPFLL} 
\end{table*}

We can construct an effective Hamiltonian for the partially empty flavors in which the interaction
is exactly the same as the one discussed in Sec.~\ref{secLLL} except that there is an 
additional two-body interaction of the screening type, in which occupied flavor electrons
are virtually excited to higher Landau levels in a completely analogous manner to $\mathcal{V}^{2b}_a(1)$ from Eq.~\eqref{V2bSLL2}, with an analogous diagramatic representation as that appearing in Fig.~\ref{figdiagrams}, except that its multiplicity will be given only by the $f_0$ occupied flavors.  This additional contribution to the effective 
interaction is:

\be\label{V2bo}
\mathcal{V}^{2b}_o =
- f_o \kappa^2 \sum_{\substack{
1\cdots6 \\ \mu\nu}
} v_{1 5 , 6 4}  v_{2 6 , 5 3} \frac{\tau(n_5,n_6)}{n_6} c^\dagger_{1 \nu} c^\dagger_{2 \mu} c_{3 \mu} c_{4 \nu},
\ee

\noindent where the orbital indices of the operators are understood to be in the $n=0$ LL, and the spin indices run over the partially empty $f-f_o$ flavors only, {\it i.e.} all electrons have minority spins when 
only spin provides a flavor label.
The additional Haldane pseudopotentials and the total Haldane pseudopotentials are listed in Table~\ref{tabVPFLL} for the most common case of spin-$1/2$ fermions.  For the case of spin-$1/2$ fermions,
only odd $m$ pseudopotentials are relevant for the Hilbert space
where this effective Hamiltonian acts.
For a more general case of $f_o$ filled flavors, the pseudopotials $V_2(m,f_o)$ are,

\be
V_2(m,f_o)=V_2(m)+f_o V_o^{2b}(m),
\ee

\noindent where the coefficients $V_2(m)$ are given by Eq.~\eqref{V2m0} and listed in Table~\ref{tabVm2}. The coefficients $V^{2b}_o(m)$, which are the pseudopotentials associated with the interaction of Eq.~\eqref{V2bo}, are listed in Table~\ref{tabVPFLL}. It is interesting to note the significant difference between the corrections to the pseudopotentials for $0<\nu<1$, listed in Table~\ref{tabVm2}, and those for spin polarized states with $1<\nu<2$, listed in Table~\ref{tabVPFLL}. For example, the correction to the $m=1$ pseudopotential, which is crucial in determining the gap of the Laughlin type states, is negative and about four times larger in magnitude compared to the $0<\nu<1$, indicating a higher reduction of this pseudopotential in the filling factor range $1<\nu<2$.

The three-body interactions between partially full flavor electrons remain unchanged and given by Eq.~\eqref{V3b}, thus, in the spin-$1/2$ case the three-body pseudopotentials would be those listed in Table~\ref{tabV3bLLL} for the $S=3/2$ states of Eq.~\eqref{Psi3/2}, constructed for the partially filled flavor with spin $\downarrow$. 

\section{Summary and Discussion}\label{secDisc} 

We have derived effective Hamiltonians which account for quantum fluctuations in full and empty Landau levels
(Landau level mixing) to leading order in perturbation theory for the cases of a partially filled $n=0$
Landau level and a partially filled $n=1$ Landau level.  These effective Hamiltonians describe 
fractional quantum Hall physics in the filling factor ranges $0 < \nu < 2$ and $2 < \nu < 4$ respectively in the case of spin-$1/2$ fermions.
In both cases the effective Hamiltonians are a sum of two- and three-body terms.
The three-body terms are responsible for  particle-hole symmetry breaking
within the Landau level. There has been considerable interest in these 
quantum fluctuation corrections because they are likely to play a decisive role in several outstanding problems in the fractional quantum Hall regime.  

Our work has been developed essentially in parallel with an analytic study by Peterson and Nayak~\cite{Peterson2013}, and a numerical study by Simon and Rezayi~\cite{Simon2013}. The three works toghether provide a comprehensive view of the leading perturbative Landau level mixing corrections to the effective Hamiltonian in the lowest and second Landau levels, and they complement earlier works by Rezayi and Haldane~\cite{Rezayi1990}, Murthy and Shankar~\cite{Murthy2002}, and Bishara and Nayak~\cite{Bishara2009}. The source of discrepancy on the two-body pseudopotentials with the work of Bishara and Nayak~\cite{Bishara2009} is now understood. In Ref.~\onlinecite{Peterson2013}, this discrepancy was shown to arise from the normal ordering that must be kept in the three body interactions. When this normal ordering is dealt with properly, our two- and three-body interactions are exactly the same as those derived in Ref.~\onlinecite{Peterson2013}. 

Our two-body pseudopotentials in the lowest and second excited Landau levels are in agreement with Refs.~\onlinecite{Peterson2013,Simon2013}. In the lowest Landau level the two-body pseudopotentials are also in agreement with the results of an earlier study of the perturbative expansion of the spectrum of hydrogenic atoms in a strong magnetic-field~\cite{MacDonald1986}. Our three-body pseudopotentials are in agreement with those of Refs.~\onlinecite{Bishara2009,Peterson2013}, and with those of Ref.~\onlinecite{Simon2013} in the lowest Landau level. There is an appreciable difference with the three-body pseudopotential differences in the second Landau level reported in Ref.~\onlinecite{Simon2013}, which is likely to arise from finite size effects in this numerical study.

Now we would like to comment on the range of validity of the perturbative approach. It is impossible to know the range of validity of any perturbative expansion without a sense for the relative size of the higher order corrections. We do not believe it is feasible to carry out the same type of analysis we have discussed in this paper to higher orders in $\kappa$ for the many-body problem in an analytic fashion. The reason is that it not possible to construct effective Hamiltonians to order $\kappa^3$ or higher, projected onto the Landau level of interest, without explicitly computing the energies to order $\kappa$.  In other words it is necessary to solve the many-body problem exactly to order $\kappa$ to be able to construct an effective Hamiltonian to order $\kappa^3$ projected into the degenerate manifold. 

In spite of this seemingly insurmountable task, it is possible to get a sense, at least heuristically, for the size of the higher order corrections. One way this can be done is by studying the two-body problem to higher order in $\kappa$. This has been done, indirectly, for the lowest Landau level in the context of the problem hydrogenic atoms in a strong magnetic field in Ref.~\onlinecite{MacDonald1986}. More specifically from the coefficients $\alpha_{N,M}^{(i)}$ listed in Table II of Ref.~\onlinecite{MacDonald1986}, one obtains the following expression for the energies of two particles with relative angular momentum $m$ in the lowest Landau level to order $\kappa^4$ in units of $\omega_c$,

\be
V_m=\sum_{p=1}^4\alpha_{0,m}^{(p)} \left(-\frac{\sqrt{\pi}}{2}\kappa\right)^p+\mathcal{O}(\kappa^5).
\ee
 
By reading the values of the coefficients $\alpha_{0,m}^{(p)}$ from Table II in Ref.~\onlinecite{MacDonald1986}, one learns that $\alpha_{0,m}^{(p)}$ decreases by about an order of magnitude at every order~\footnote{This decrease is not that fast for the $m=0$ state, but it is very fast for $m\geq1$.}. This indicates that higher order corrections remain parametrically small even at $\kappa \sim1$. This observation suggests that higher order corrections might remain small even at values of $\kappa \sim1$ in the Lowest landau level, and specially so in the dilute limit where the energies of the two-body problem are expected to dominate. 

It is not possible to directly extract the energies of two fermions in the second landau level togheter with the completely filled lowest Landau level from the results of the bare two-body problem of Ref.~\onlinecite{MacDonald1986}. This is because the bare two-body problem fails to account for basic many-body effects like Pauli blocking that already arise at order $\kappa$ in the energies in the second Landau level. It is thus hard to anticipate at this point the relative size of the $\kappa^3$ contributions in the second Landau level.

Most fractional quantum Hall samples have similar electron densities.  For this reason the external magnetic field strength tends to be smaller, and $\kappa$ correspondingly larger, for the experiments in the second Landau level than experiments in the lowest Landau level. Additionally by comparing Tables~\ref{tabVm2} and~\ref{tabVm2SLL} it is evident that even at a fixed field, the two-body pseudopotential 
corrections are larger for the second Landau level than for the lowest Landau level. Quantum fluctuations in Landau level occupations are therefore more likely to be important in the second Landau level case in which the fractional quantum Hall effect can be enriched by the appearance of striped states and even-denominator incompressible states.

Finally we would like to connect our study to the problem of the nature of the incompressible state observed at filling fraction $\nu=5/2$ in GaAs. For the spin polarized case, the Moore-Read Pfaffian is known to be the unique highest density zero-energy state of a repulsive three-body Hamiltonian~\cite{Greiter1991,Greiter1992,Wen1993,Rezayi2000} for which only the lowest angular momentum three-body state is energetically penalized, namely $V_{3/2}(m=3)>0$ and the remainder pseudopotentials vanish~\cite{Simon2007}. Conversely the anti-Pfafffian is expected to be the ground for the particle-hole conjugated Hamiltonian~\cite{Wang2009}, which has $V_{3/2}(m=3)<0$. We have found in agreement with Refs.~\onlinecite{Bishara2009,Peterson2013} that the leading value of this 
pseudopotential is negative. 
Nevertheless the two-particle pseudopotentials, which we have corrected for in this work, have been found to play decisive role in this competition, and they could drive the system into a compressible phase.~\cite{Morf1998,Rezayi2000,Moller2008,Peterson2008b,Peterson2008,Peterson2008a,Wojs2009,Wang2009,Storni2010,Lu2010,Wojs2010,Wojs2010a,Rezayi2011,Papic2012}
A reliable assessment of the influence of LL mixing on fractional quantum Hall states at even denominator fractions thus awaits 
the application of our pseudopotentials in many-body exact diagonalization studies.  

\begin{acknowledgements}

We are grateful to Michael Peterson, Chetan Nayak, Steven Simon and Edward Rezayi for valuable discussions and correspondence, and for sharing unpublished information which helped us identify an error in an earlier unpublished draft of this Manuscript. We would like to thank Zlatko Papic for sharing his unpublished results and for spotting an issue in our three-body pseudopotential matrix in the second Landau level which we have now corrected. We are thankful as well to Thierry Joelicour, Arkadiusz Wojs, and Ramamurti Shankar for valuable correspondence and comments on an earlier unpublished version of this manuscript.

This work was supported by the Welch foundation under grant TBF1473, by the DOE Division of Materials Sciences and Engineering under grant DE-FG03-02ER45958, and by the NRI SWAN program. 

\end{acknowledgements}

\appendix

\section{Three-body states and pseudopotentials}\label{3bapp}

In this appendix we outline some properties of the three body states and their associated generalized Haldane pseudopotentials. 
We begin with the $n=0$ LL. We are interested only in translationally invariant polynomials describing the internal state of relative motion of the three particles, and thus omit polynomial factors in the center of mass coordinates $(z_1+z_2+z_3)/3$. Translationally invariant polynomials would depend only on two independent translationally invariant coordinates, which can be chosen as

\be
z_a=\sqrt{\frac{2}{3}}\left(\frac{z_1+z_2}{2}-z_3\right), \ \ z_b=\frac{z_1-z_2}{\sqrt{2}}.
\ee 
 
The permutation operators on these coordinates act as a reflection, and as a composition of a reflection and rotations by $\pm 2\pi/3$~\cite{Laughlin1983}, 

\be\label{permmat}
\Pi_{12}
\dot{=}
\begin{pmatrix}
1& 0\\
0& -1
\end{pmatrix},  \
\Pi_{23}
\dot{=}
\begin{pmatrix}
-\frac{1}{2}& \frac{\sqrt{3}}{2}\\
\frac{\sqrt{3}}{2}& \frac{1}{2}
\end{pmatrix},  \
\Pi_{13}
\dot{=}
\begin{pmatrix}
-\frac{1}{2}& -\frac{\sqrt{3}}{2}\\
-\frac{\sqrt{3}}{2}& \frac{1}{2}
\end{pmatrix}, 
\ee

\noindent where the matrices are understood to act in a colum vector of the form $\bigl( \begin{smallmatrix}z_a\\z_b\end{smallmatrix} \bigr)$.

The fully spin polarized $S=3/2$ states, whose polynomial part is fully antisymmetric, were constructed in Ref.~\onlinecite{Laughlin1983}, and read as

\be\label{Psi3/2app}
\Psi^{\scriptscriptstyle{3/2}}_{kl}=\frac{1}{Z^{\scriptscriptstyle{3/2}}_{kl}}(z_a^2+z_b^2)^{k}\left[\frac{(z_a+iz_b)^{3 l}-(z_a-iz_b)^{3 l}}{2i}\right],
\ee

\noindent with $Z^{\scriptscriptstyle{3/2}}_{kl}=2^{3l+2k+1}[\pi^3(3l+k)! k!]^{1/2}$, $k\geq0$, $l\geq1$. $\Psi^{\scriptscriptstyle{3/2}}_{kl}$ can be expanded as,

\be
\Psi^{\scriptscriptstyle{3/2}}_{kl}=\frac{1}{Z^{\scriptscriptstyle{3/2}}_{kl}}\sum_{j=0}^{m}f^{\scriptscriptstyle{3/2}}_{jkl} \ z_b^j z_a^{m-j}
\ee

\noindent where $m=2k+3l$ is the orbital angular momentum of $\Psi^{\scriptscriptstyle{3/2}}_{kl}$, and

\be
f^{\scriptscriptstyle{3/2}}_{jkl}=\sum_{p=0}^{k}\sum_{q=0}^{3l}\binom{k}{p}\binom{3l}{q}\sin(q\pi/2)\delta_{j,q+2p}.
\ee

In order to evaluate the pseudopotentials from Eq.~\eqref{spin3pseudo} it is convenient to decompose the state $\Psi^{\scriptscriptstyle{3/2}}_{kl}$ into products of states with well defined relative numbers for the pair of particles $1$ and $2$, and states with well defined numbers for particle $3$, as follows:

\be\label{3/2exp}
\Psi^{\scriptscriptstyle{3/2}}_{kl}=\sum_{j=0}^m\sum_{j'=0}^{m-j} C_{kl,jj'}|m-j-j',j\rangle_{12}|j'\rangle_3,
\ee

\noindent where $|m,m'\rangle_{ij}$ abbreviates for the state with center of mass angular momentum $m$ and relative angular momentum $m'$ for particles $i$ and $j$ in the $n=0$ LL, and $|l\rangle_i$ for the state of particle $i$ with angular momentum $l$ in the $n=0$ LL. $C_{kl,jj'}$ can be found to be:

\begin{multline}
C_{kl,jj'}=(2\pi)^{3/2}\frac{f^{\scriptscriptstyle{3/2}}_{jkl}}{Z^{\scriptscriptstyle{3/2}}_{kl}}\binom{m-j}{j'}(-1)^{j'}\\
\times\sqrt{\frac{2^{m+j'}}{3^{m-j}}(m-j-j')!j!j'!}.
\end{multline}

Consider now two states, $\Psi^{\scriptscriptstyle{3/2}}_{kl}$ and $\Psi^{\scriptscriptstyle{3/2}}_{k'l'}$, with the same angular momentum, $m=2k+3=2k'+3l'$. Their associated generalized Haldane pseudopotential matrix elements, computed from Eqs.~\eqref{spin3pseudo} and~\eqref{V3b3/2}, is

\begin{widetext}
\be
\label{V3b3/2full}
V_{3/2}(k'l',kl)=6\sum_{i=0}^m\sum_{i'=0}^{m-i} C_{k'l',ii'}\sum_{j=0}^m\sum_{j'=0}^{m-j}C_{kl,jj'}\sum_{n=1}^\infty\frac{{}_r\langle 0,i| v_{12} |n,i+n\rangle_r{}_r\langle n,j+n| v_{12} |0,j\rangle_r}{2^n n}R^{m+n-i'}_{i+n,j'}R^{m+n-j'}_{j+n,i'},
\ee
\end{widetext}

\noindent where $R^L_{m,m'}$ are given by Eq.~\eqref{Rmat}, and the Coulomb matrix elements by Eq.~\eqref{coulmatrix}. Equation~\eqref{V3b3/2full} can be used to obtain the $S=3/2$ pseudopotentials listed in Table~\ref{tabV3bLLL}. 

Let us now discuss the $S=1/2$ three-body states. We begin by considering the bosonic analogue of $\Psi^{\scriptscriptstyle{3/2}}_{kl}$, namely

\be
\Phi_{kl}=(z_a^2+z_b^2)^{k}\left[\frac{(z_a+iz_b)^{3 l}+(z_a-iz_b)^{3 l}}{2}\right],
\ee

\noindent with $k\geq0$, $l\geq0$, and we have not normalized these states yet. These states form a complete orthogonal basis for the fully symmetric translationally invariant polynomials. They can be expanded as

\be
\Phi_{kl}=\sum_{j=0}^{2k+3l}f^{\scriptscriptstyle{1/2}}_{jkl} \ z_b^j z_a^{2k+3l-j},
\ee

\noindent with

\be
f^{\scriptscriptstyle{1/2}}_{jkl}=\sum_{p=0}^{k}\sum_{q=0}^{3l}\binom{k}{p}\binom{3l}{q}\cos(q\pi/2)\delta_{j,q+2p}.
\ee

The spatial part of the $S=1/2$ states can the be written as~\cite{Davenport2012}

\be
\Psi^{\scriptscriptstyle{1/2}}_{\sigma kl}=\frac{\beta_\sigma}{Z^{\scriptscriptstyle{1/2}}_{\sigma kl}}\Phi_{kl},
\ee

\noindent with $\sigma=\{1,2\}$, $k\geq0$, $l\geq0$, with the normalization constants $Z^{\scriptscriptstyle{1/2}}_{\sigma kl}$ given in the text below Eq.~\eqref{S=3/2Psi}. Where the polynomials $\beta_\sigma$ are~\footnote{In the notation of Ref.~\onlinecite{Davenport2012}, $\beta_1=\sqrt{2} z_b$ and $\beta_2=2 z_a z_b/\sqrt{3}$.}

\be
\beta_1=z_b, \ \beta_2=z_az_b.
\ee

This basis is complete but not orthogonal. Fortunately, the inner products can be computed analytically and are listed in Eq.~\eqref{innerproducts}, thus othonormalization becomes trivial. 

The fully antisymmetric wavefunction including the spin part is~\cite{Davenport2012}

\be\label{antisym}
|\Psi^{\scriptscriptstyle{1/2}}_{\sigma kl}\rangle=\mathcal{A}\{\Psi^{\scriptscriptstyle{1/2}}_{\sigma kl}\otimes |\uparrow\uparrow\downarrow\rangle\},
\ee

\noindent where $\mathcal{A}=(1-\Pi_{23}+\Pi_{12}\Pi_{23})/\sqrt{3}$ is the 
antisymmetrization operator between $\uparrow$ and $\downarrow$ particles, 
and $\Pi_{ij}$ is the unitary operator corresponding to elementary permutations between $i$ and $j$. 

The three-body interaction in Eq.~\eqref{V3b} is rotationally invariant, hence, its associated three-body pseudopotentials are diagonal in the relative orbital angular momentum of $|\Psi^{\scriptscriptstyle{1/2}}_{\sigma kl}\rangle$, $m=\sigma+2k+3l$, therefore, for any two states with the same angular momentum we can write

\begin{multline}\label{V3b1/2app}
V_{1/2}(\sigma'k'l',\sigma kl)=-6 \langle\Psi^{\scriptscriptstyle{1/2}}_{\sigma' k'l'}|v_{13}\mathcal{P}_{\perp} \frac{1}{\hat{n}} \mathcal{P}_{\perp}v_{12}|\Psi^{\scriptscriptstyle{1/2}}_{\sigma kl}\rangle\\
=
2 \bar{\Psi}^{\scriptscriptstyle{1/2}}_{\sigma' k'l'}(h_0+h_0^\dagger+h_1)\Psi^{\scriptscriptstyle{1/2}}_{\sigma kl},
\end{multline}

\noindent where $\hat{n}=\hat{n}_1+\hat{n}_2+\hat{n}_3$, $\mathcal{P}_{\perp}$ is the projector into the othogonal complement to the $\hat{n}=0$ eigenspace. In the second line of Eq.~\eqref{V3b1/2app} it is understood that only the orbital part of the wavefunctions is involved, and it follows from Eq.~\eqref{antisym} using the fact that interactions are spin independent. The computation of Haldane pseudopotentials reduces to the computation of the matrix elements of $h_0$ and $h_1$, which stand for

\be
h_0=-v_{13}\mathcal{P}_{\perp} \frac{1}{\hat{n}} \mathcal{P}_{\perp}v_{12}, \ \
h_1=-v_{13}\mathcal{P}_{\perp} \frac{1}{\hat{n}} \mathcal{P}_{\perp}v_{23}.
\ee

In analogy with Eq.~\eqref{3/2exp}, we can decompose $\Psi^{\scriptscriptstyle{1/2}}_{\sigma kl}$ as

\be\label{1/2exp}
\Psi^{\scriptscriptstyle{1/2}}_{\sigma kl}=\sum_{j=0}^m\sum_{j'=0}^{m-j} C^{\mu\nu}_{\sigma kl,jj'}|m-j-j',j\rangle_{\mu\nu}|j'\rangle_\gamma,
\ee

\noindent where $m=\sigma+2k+3l$, and the indices $\{\mu\nu\gamma\}$ stand for any permutation of $\{123\}$. The decomposition is generally dependent on the particle ordering since the orbital part alone of the $S=1/2$ states, $\Psi^{\scriptscriptstyle{1/2}}_{\sigma kl}$, is not an eigenstate of all permutations. $C^{\mu\nu}_{\sigma kl,jj'}$ can be found to be

\begin{multline}
C^{\mu\nu}_{\sigma kl,jj'}=(2\pi)^{3/2}\frac{g^{\mu\nu}_{\sigma jkl}}{Z^{\scriptscriptstyle{1/2}}_{\sigma kl}}\binom{m-j}{j'}(-1)^{j'}\\
\times\sqrt{\frac{2^{m+j'}}{3^{m-j}}(m-j-j')!j!j'!},
\end{multline}

\noindent with

\be
\begin{split}
g^{12}_{\sigma jkl}&=f^{\scriptscriptstyle{1/2}}_{j-1,kl}, \\ 
g^{13}_{1 jkl}&=\frac{f^{\scriptscriptstyle{1/2}}_{j-1,kl}+\sqrt{3}f^{\scriptscriptstyle{1/2}}_{j,kl}}{2},\\
g^{32}_{1 jkl}&=\frac{f^{\scriptscriptstyle{1/2}}_{j-1,kl}-\sqrt{3}f^{\scriptscriptstyle{1/2}}_{j,kl}}{2}, \\
g^{13}_{2 jkl}&=\frac{2 f^{\scriptscriptstyle{1/2}}_{j-1,kl}+\sqrt{3}(f^{\scriptscriptstyle{1/2}}_{j-2,kl}-f^{\scriptscriptstyle{1/2}}_{j,kl})}{4},\\
g^{32}_{2 jkl}&=\frac{2 f^{\scriptscriptstyle{1/2}}_{j-1,kl}-\sqrt{3}(f^{\scriptscriptstyle{1/2}}_{j-2,kl}-f^{\scriptscriptstyle{1/2}}_{j,kl})}{4}.
\end{split}
\ee

\noindent These expressions need not be derived independently, but can be obtained by deriving only the decomposition corresponding to $\{\mu\nu\}=\{12\}$, and then applying suitable permutation operators as represented in Eq.~\eqref{permmat}. With this the evaluation of Eq.~\eqref{V3b1/2app} leads to,  

\begin{widetext}
\begin{multline}
\label{V3b3/2full}
\bar{\Psi}^{\scriptscriptstyle{1/2}}_{\sigma' k'l'}h_0\Psi^{\scriptscriptstyle{1/2}}_{\sigma kl}=-\sum_{i=0}^m\sum_{i'=0}^{m-i} C^{13}_{\sigma'k'l',ii'}\sum_{j=0}^m\sum_{j'=0}^{m-j}C^{12}_{\sigma kl,jj'}\sum_{n=1}^\infty\frac{{}_r\langle 0,i| v_{12} |n,i+n\rangle_r{}_r\langle n,j+n| v_{12} |0,j\rangle_r}{2^n n}R^{m+n-i'}_{i+n,j'}R^{m+n-j'}_{j+n,i'},\\
\bar{\Psi}^{\scriptscriptstyle{1/2}}_{\sigma' k'l'}h_1\Psi^{\scriptscriptstyle{1/2}}_{\sigma kl}=-\sum_{i=0}^m\sum_{i'=0}^{m-i} (-1)^iC^{13}_{\sigma'k'l',ii'}\sum_{j=0}^m\sum_{j'=0}^{m-j}C^{32}_{\sigma kl,jj'}\sum_{n=1}^\infty\frac{{}_r\langle 0,i| v_{12} |n,i+n\rangle_r{}_r\langle n,j+n| v_{12} |0,j\rangle_r}{2^n n}R^{m+n-i'}_{i+n,j'}R^{m+n-j'}_{j+n,i'}.
\end{multline}
\end{widetext}

\noindent Notice that all the matrix elements are purely real. Combining these expressions with Eqs.~\eqref{coulmatrix},~\eqref{Rmat} and~\eqref{V3b1/2app}, one obtains the values listed in Table~\ref{tabV3bLLL}. 

The derivation in the $n=1$ LL goes through in a completely analogous fashion. By raising the kinetic energy of the three particles, with the operator $a_1^\dagger a_2^\dagger a_3^\dagger$, we easily obtain the required representation of the states in one-to-one correspondence with the $n=0$ LL. 

To obtain the $S=3/2$ three-body pseudopotentials in the $n=1$ LL we apply  $a_1^\dagger a_2^\dagger a_3^\dagger$ to Eq.~\eqref{3/2exp}. Using the first quantized version of the three-body interactions appearing in Eq.~\eqref{V3bSLL2}, we arrive at the following expression for the pseudopotentials in the $n=1$ LL, 

\begin{widetext}
\begin{multline}\label{S=3/2SLLpseudo}
V_{3/2}(k'l',kl)=6\sum_{i=0}^m\sum_{i'=0}^{m-i} C_{k'l',ii'}\sum_{j=0}^m\sum_{j'=0}^{m-j}C_{kl,jj'}\biggl[\sum_{n=1}^\infty\frac{{}_r( 0,i| v_{12} |n,i+n)_r{}_r( 0,j| v_{12} |n,j+n)_r}{2^n n} R^{m+n-i'}_{i+n,j'}R^{m+n-j'}_{j+n,i'}\\
-\frac{{}_r\langle 2,i| v_{12} |1,i-1\rangle_r {}_r\langle 1,j-1| v_{12} |2,j\rangle_r}{4} R^{m-1-i'}_{i-1,j'}R^{m-1-j'}_{j-1,i'}\biggr],
\end{multline}
\end{widetext}

\noindent where we have introduced the notation,

\begin{multline}
{}_r( 0,i| v_{12} |n,i+n)_r\equiv {}_r\langle 0,i| v_{12} |n,i+n\rangle_rR^{n+2}_{n,1}\\
-{}_r\langle 2,i| v_{12} |n+2,i+n\rangle_rR^{n+2}_{n+2,1}.
\end{multline}

From Eq.~\eqref{S=3/2SLLpseudo} one obtains the pseudopotentials for $S=3/2$ states in the $n=1$ LL listed in Table~\ref{tabV1/2SLL}.

To obtain the $S=1/2$ states in the $n=1$ LL we apply  $a_1^\dagger a_2^\dagger a_3^\dagger$ to Eq.~\eqref{1/2exp}. From the interaction appearing in Eq.~\eqref{V3bSLL2} we obtain the analogue of Eq.~\eqref{V3b1/2app}, now with $h_0$ and $h_1$ replaced by

\be
\tilde{h}_0=-v_{13}\mathcal{P}_{\perp} \frac{1}{\hat{n}-3} \mathcal{P}_{\perp}v_{12}, \ \
\tilde{h}_1=-v_{13}\mathcal{P}_{\perp} \frac{1}{\hat{n}-3} \mathcal{P}_{\perp}v_{23}.
\ee

\noindent The matrix elements for $\tilde{h}_0$ are,

\begin{widetext}
\begin{multline}
\label{V3b1/2full}
\bar{\Psi}^{\scriptscriptstyle{1/2}}_{\sigma' k'l'}\tilde{h}_0\Psi^{\scriptscriptstyle{1/2}}_{\sigma kl}=-\sum_{i=0}^m\sum_{i'=0}^{m-i} C^{13}_{\sigma'k'l',ii'}\sum_{j=0}^m\sum_{j'=0}^{m-j}C^{12}_{\sigma kl,jj'}\biggl[\sum_{n=1}^\infty\frac{{}_r( 0,i| v_{12} |n,i+n)_r{}_r( 0,j| v_{12} |n,j+n)_r}{2^n n} R^{m+n-i'}_{i+n,j'}R^{m+n-j'}_{j+n,i'}
\\
-\frac{{}_r\langle 2,i| v_{12} |1,i-1\rangle_r {}_r\langle 1,j-1| v_{12} |2,j\rangle_r}{4} R^{m-1-i'}_{i-1,j'}R^{m-1-j'}_{j-1,i'}\biggr].
\end{multline}
\end{widetext}

The expression for $\bar{\Psi}^{\scriptscriptstyle{1/2}}_{\sigma' k'l'}\tilde{h}_1\Psi^{\scriptscriptstyle{1/2}}_{\sigma kl}$ is the same after replacing $C^{12}_{\sigma kl,jj'}\rightarrow C^{32}_{\sigma kl,jj'}$, and $C^{13}_{\sigma' k'l',ii'}\rightarrow (-1)^iC^{13}_{\sigma' k'l',ii'}$, in analogy with Eq.~\eqref{V3b3/2full} for the $n=0$ LL. The $S=1/2$ Haldane pseudopotentials in the $n=1$ LL, listed in Table~\ref{tabV1/2SLL}, are given $V_{1/2}(\sigma'k'l',\sigma kl)=2 \bar{\Psi}^{\scriptscriptstyle{1/2}}_{\sigma' k'l'}(\tilde{h}_0+\tilde{h}_0^\dagger+\tilde{h}_1)\Psi^{\scriptscriptstyle{1/2}}_{\sigma kl}$, which is the analogue of Eq.~\eqref{V3b1/2app} for the $n=0$ LL.

In spite of how cumbersome Eqs.~\eqref{S=3/2SLLpseudo} and~\eqref{V3b1/2full} might look, its evaluation is very efficient, and each pseudopotential takes only a few seconds to evaluate using {\it Mathematica} in a conventional laptop computer.

We listed in Tables~\ref{tabV3bLLL} and~\ref{tabV1/2SLL} pseudopotentials matrices up to $2\times2$ in size. For this case the two-dimensional subspaces for given $m$ contain one state for $\sigma=1$ and one for $\sigma=2$. We have orthogonalized these states by rotating the $\sigma=2$ state only. Let us call the matrix in the non-orthogonal basis ${\bf V}$, which is obtained from Eqs.~\eqref{V3b1/2app} and~\eqref{V3b3/2full}. The matrix listed in Tables~\ref{tabV3bLLL} and~\ref{tabV1/2SLL} corresponds to $({\bf B}^T)^{-1} {\bf V} {\bf B}^{-1}$, with ${\bf B}$ the change of basis matrix,

\be
{\bf B}=\begin{pmatrix}
1& \langle\Psi_1|\Psi_2\rangle\\
0 & \sqrt{1-\langle\Psi_1|\Psi_2\rangle^2}
\end{pmatrix}.
\ee  

\noindent $\langle\Psi_1|\Psi_2\rangle$ is a shorthand for the overlap between $\sigma=\{1,2\}$ states, appearing in Eq.~\eqref{innerproducts}.

\section{Effective interactions in the $n=1$ LL}\label{intSLL}

In this appendix we describe in more detail how the interactions in the $n=1$ LL were obtained. Consider the effective Hamiltonian to order $\kappa^2$ in the $n=1$ LL,

\begin{multline}\label{H2SLLApp}
\mathcal{H}_2=\mathcal{H}_1-\frac{\kappa^2}{4}  \sum_{\substack{
1\cdots8 \\
\mu\nu\lambda\sigma}} v_{1 2 , 3 4}  v_{5 6 , 7 8}\\
\times P_0 c^\dagger_{5 \nu} c^\dagger_{6 \mu} c_{7 \mu} c_{8 \nu} P_\perp \frac{1}{\hat{n}-N_1} P_\perp c^\dagger_{1 \lambda} c^\dagger_{2 \sigma} c_{3 \sigma} c_{4 \lambda} P_0,
\end{multline}

\noindent where the integers abbreviate single-particle kinetic and guiding center quantum numbers ({\it e.g.} $\{1 \Leftrightarrow n_1m_1 \}$),  $N_1$ is the number of particles in the 
partially filled $n=1$ LL, $P_0$ is the projector into the many-body eigenspace with $\hat{n}=N_1$, and $P_\perp$ 
the projector into its orthogonal complement. 

The second order term in Eq.~\eqref{H2SLLApp} can be viewed as a sequence of scattering process in which a pair of particles is taken from states ``3" and ``4", contained either in the completely filled $n=0$ LL or the partially filled $n=1$ LL, and placed into states ``1" and ``2" with higher total kinetic energy. Subsequently, the particles are removed from these virtually excited states by operators ``7" and ``8", to be finally placed back into states ``5" and ``6", which are contained either in the $n=0$ LL or the $n=1$ LL. 

Threrefore any term with non-vanishing matrix elements in the $\hat{n}=N_1$ eigenspace must have the outermost destruction and creation operators ({\it i.e.} $c^\dagger_{5 \nu}$, $c^\dagger_{6 \mu}$,  $c_{3 \sigma}$ and $c_{4 \lambda}$) with Landau level indices either $0$ or $1$. Therefore, the kinetic energy of the ``incoming" particles, {\it i.e.} $n_3+n_4$, or the kinetic energy of the ``outgoing" particles, {\it i.e.} $n_5+n_6$, is allowed to take only the values $\{0,1,2\}$. It can be shown that there is a kinetic energy balance condition between ``incoming" and ``outgoing" labels, namely the terms with $n_3+n_4\neq n_5+n_6$ vanish, and they correspond to scattering type diagrams that have tad-pole and self-energy insertions in their legs. In summary, we can label the allowed interactions by the total incoming/outgoing pair kinetic energy $n_3+n_4= n_5+n_6=\{0,1,2\}$. 

As an example we discuss in detail how to obtain the interaction corresponding to incoming pairs with total kinetic energy $n_3+n_4=2$, labeled $\mathcal{V}^{2b}(2)$, and will leave the verification of the remainding terms to the interested reader. This term corresponds to all the two-body terms arising from Eq.~\eqref{H2SLLApp} with $n_3=n_4=n_5=n_6=1$, after normal ordering of the operators is performed. In order to have a non-vanising contribution the creation operators $c^\dagger_{1 \lambda}$ and $c^\dagger_{2 \sigma}$ must raise the kinetic energy of the incoming pair, thus $n_1+n_2>2$.  
Similarly we have that $n_7+n_8>2$. Since the $n=0$ LL is assumed to be completely full, these labels are additionally constrained to satisfy $n_i\geq1$, for $i=\{1,2,7,8\}$. 

With these constraints we can see that there are two possibilities for these middle operators.  Either neither of them corresponds to the partially filled $n=1$ LL, {\it i.e.} $n_i>1$, or else only one creation and only one destruction do correspond to $n=1$ LL, for example $n_2=n_7=1$ and $n_1=n_8>1$. For the first possibility we see that the indices $\{1,2,7,8\}$ must be fully contracted in order for the virtually excited pair to go back into the $n=1$ LL.
Let us name the terms arising from this first possibility as $\mathcal{V}^{2b}_a(2)$.  We can write

\begin{multline}\label{H2SLLApp2}
\mathcal{V}^{2b}_a(2)=-\frac{\kappa^2}{4}  \sum_{\substack{
1\cdots8 \\
\mu\nu\lambda\sigma}} v_{1 2 , 3 4}  v_{5 6 , 7 8} \langle c_{7 \mu} c_{8 \nu} c^\dagger_{1 \lambda} c^\dagger_{2 \sigma}\rangle \\ \times\frac{\vartheta(n_1)\vartheta(n_2)}{n_1+n_2-2} c^\dagger_{5 \nu} c^\dagger_{6 \mu} c_{3 \sigma} c_{4 \lambda},
\end{multline}

\noindent where now we understand the Landau level index of the uncontracted operators to be in the $n=1$ LL, and $\vartheta(n)$ restricts $n\geq2$. 

For the second possibility an extra pair of uncontracted creation and destruction operators in Eq.\eqref{H2SLLApp} will have indices in the $n=1$ LL.  Let us name this term $\mathcal{V}_b(2)$, which reads as

\begin{multline}\label{H2SLLApp3}
\mathcal{V}_b(2)=-\frac{\kappa^2}{4} \sum_{\substack{
1\cdots8 \\
\mu\nu\lambda\sigma}} v_{1 2 , 3 4}  v_{5 6 , 7 8} c^\dagger_{5 \nu} c^\dagger_{6 \mu} \Bigl[\frac{\vartheta(n_1)}{n_1-1} \\
\times\Bigl(\langle c_{8 \nu} c^\dagger_{1 \lambda}\rangle c_{7 \mu}  c^\dagger_{2 \sigma}-\langle c_{7 \mu} c^\dagger_{1 \lambda}\rangle c_{8 \nu} c^\dagger_{2 \sigma}\Bigl) +\{1  \Leftrightarrow 2\}\Bigr]  c_{3 \sigma} c_{4 \lambda},
\end{multline}

\noindent where again we understand the Landau level index of the uncontracted operators to be in the $n=1$ LL. 

It is clear from Eq.~\eqref{H2SLLApp3}, that $\mathcal{V}_b(2)$ has a contribution to the effective three body interactions, but it also contributes to the two body interactions after due normal ordering of the operators is performed. The three body piece in Eq.~\eqref{H2SLLApp3} contributes to the terms in Eq.~\eqref{V3bSLL}, with $n_7\geq2$. On the other hand, the two body term, which we label $\mathcal{V}^{2b}_b(2)$, makes up the remainder of the interaction $\mathcal{V}^{2b}(2)$ appearing in Eq.~\eqref{V2bSLL}, namely $\mathcal{V}^{2b}(2)=\mathcal{V}^{2b}_a(2)+\mathcal{V}^{2b}_b(2)$. 

A very similar analysis 
leads to the forms for the remainding two- and three-body interactions given in Eqs.~\eqref{V2bSLL},~\eqref{V2bSLL2},~\eqref{V3bSLL}, and~\eqref{V2bo}.  

\section{Two-body states and Haldane pseudopotentials in the $n=1$ LL}\label{2bpseud} 

In this section we discuss several useful properties of the two-body states in the $n=1$ LL and their associated Haldane pseudopotentials. 

We begin by discussing two-particle states. Consider a state for particles $i$ and $j$ in which each of them has well defined kinetic and guiding center quantum numbers: 

\be\label{12basis}
|n_1 m_1\rangle_i|n_2 m_2\rangle_j= \frac{a_i^{\dagger n_1}a_j^{\dagger n_2}b_i^{\dagger m_1}b_j^{\dagger m_2}}{\sqrt{n_1!n_2!m_1!m_2!}}|0\rangle,
\ee

\noindent where $a^{\dagger}$ and $b^{\dagger}$ are the kinetic energy and guiding center raising operators~\cite{cond-mat/9410047,Giuliani2005}.  Although we often loosely refer to the guiding center quantum number $m_i$ as 
the angular momentum of particle $i$, notice that the physical angular momentum is actually $m_i-n_i$. 

An alternative representation for two-particle states is obtained by constructing 
states with well defined center of mass and relative coordinate quantum numbers, for both the kinetic energy and guiding center labels, as follows

\be\label{CMrbasis}
|N M\rangle_c|n m\rangle_r= \frac{A^{\dagger N}B^{\dagger M} a^{\dagger n}b^{\dagger m}}{\sqrt{N!M!n!m!}}|0\rangle,
\ee

\noindent where the four operators $A=(a_i+a_j)/\sqrt{2}$, $B=(b_i+b_j)/\sqrt{2}$, $a=(a_i-a_j)/\sqrt{2}$ and $b=(b_i-b_j)/\sqrt{2}$, commute with each other. These states are related to those in Eq.~\eqref{12basis} by the unitary transformation

\begin{multline}\label{transf}
|n_1 m_1\rangle_i|n_2 m_2\rangle_j=\sum_{\nu=0}^{n_1+n_2} \sum_{\mu=0}^{m_1+m_2} R^{n_1+n_2}_{n_2,\nu} R^{m_1+m_2}_{m_2,\mu}\\
\times |n_1+n_2-\nu, m_1+m_2-\mu\rangle_{c}|\nu \mu\rangle_r, 
\end{multline}

\noindent where $R^{L}_{m,m'}$ is a real, orthogonal, and symmetric $L\times L$ matrix. The explicit form of $R^{L}_{m,m'}$ is

\begin{multline}\label{Rmat}
R^L_{m,m'}=
\sqrt{\frac{\binom{L}{m}}{2^L\binom{L}{m'}}}\\
\times\sum_{\mu=Max(0,m+m'-L)}^{min(m',m)} \binom{L-m}{m'-\mu}\binom{m}{\mu} (-1)^\mu.
\end{multline}

\noindent The properties of $R^{L}_{m,m'}$ imply that the inverse transformation is still given by Eq.~\eqref{12basis} with the exchange of roles $\{i\Leftrightarrow c, j\Leftrightarrow r\}$. 

For two isolated particles in vacuum the center of mass and relative basis of Eq.~\eqref{CMrbasis} is the most natural basis to find eigenstates of the two-body problem in the presence of strong magnetic fields. The reason is that the interaction is almost diagonal in this basis:

\begin{multline}
{}_c\langle N' M'|{}_r\langle n' m'|v_{ij}|N M\rangle_c|n m\rangle_r\\
={}_r\langle n' m'|v_{ij}|n m\rangle_r \delta_{m-n,m'-n'} \delta_{N,N'}\delta_{M,M'}.
\end{multline}

For the Coulomb potential, $v_{ij}=1/|r_i-r_j|$, with lengths measured in $l_0$ units, the non-vanishing matrix elements are:

\begin{multline}\label{coulmatrix}
{}_r\langle n' m'|v_{ij}|n m\rangle_r=\frac{\Gamma(|j|+1/2)\Gamma(l'+1/2)}{2 |j|!}\\ \times\sqrt{\frac{(l+|j|)!}{\pi(l'+|j|)!l'!l!}}\ \pFq{3}{2}{-l,|j|+\frac{1}{2},\frac{1}{2}}{|j|+1,1/2-l'}{1},
\end{multline}

\noindent where $m\geq0$, $m'\geq0$, $n\geq0$, $n'\geq0$, $m'-n'=m-n$, $j=m-n$, $l=n+(j-|j|)/2$, and $l'=n'+(j-|j|)/2$. Equation~\eqref{coulmatrix} can be obtained by combining the explicit expressions for the relative coordinate wavefunctions~\cite{Giuliani2005} with the useful integrals computed in the Appendix of Ref.~\onlinecite{Glasser1985}. 

For the $N=f N_{LL} +2$-body problem, in which the lowest Landau level is completely filled and only two particles are in the $n=1$ LL, the basis of pure relative and center of mass quantum numbers is no longer the most convenient choice to describe the relative motion of these two particles. The reason is that generally states with well defined relative kinetic energy, $n>1$, will have a probability amplitude for the individual particles to occupy the $n=0$ LL, which is forbidden by the Pauli exclusion principle. The most natural states would look entangled in both the basis of Eq.~\eqref{12basis} and Eq.~\eqref{CMrbasis}, and have the form

\begin{multline}\label{entangled}
|M,m\rangle_1\equiv a^\dagger_1 a^\dagger_2 |M,0\rangle_c|m,0\rangle_r\\
=\frac{1}{\sqrt{2}}(|M,2\rangle_c|m,0\rangle_r-|M,0\rangle_c|m,2\rangle_r).
\end{multline}

\noindent These are the states that determine the Haldane pseudopotentials in the $n=1$ LL~\cite{Haldane1987,Simon2007} appearing in Eq.~\eqref{V1SLL}. To simplify notation, we have denoted them simply in Eq.~\eqref{V1SLL} by $|m\rangle_1$, since the pseudopotentials are diagonal on and independent of the center of mass label $M$. 

The computation of Haldane pseudopotentials in the $n=1$ LL is simplified by transforming the two body interactions of Eqs.~\eqref{V2bSLL},~\eqref{V2bSLL2},~\eqref{V3bSLL}, and~\eqref{V2bo}, into their first quantization version, which we describe in the  remainder of this Appendix. For $\mathcal{V}^{2b}(2)$ appearing in Eq.~\eqref{V2bSLL}, we have the equivalent first quantization form

\be\label{fstQV2}
\mathcal{V}^{2b}_{ij}(2)=-\kappa^2v_{ij}\biggl[\sum_{1,2}|1_i 2_j\rangle \langle1_i 2_j| \frac{\theta(n_1,n_2)}{n_1+n_2-2}\biggr]v_{ij},
\ee   

\noindent where $|1_i 2_j\rangle$ is a shorthand for the state $|n_1 m_1\rangle_i|n_2 m_2\rangle_j$, and
$\theta(n_1,n_2)$ restricts $n_1+n_2\geq3$ and $n_1\geq1,n_2\geq1$. Combining Eqs.~\eqref{transf},~\eqref{Rmat}, and~\eqref{entangled} with Eq.~\eqref{fstQV2} leads to the following semi-analytic expression for the Haldane pseudopotentials associated with $\mathcal{V}^{2b}(2)$:

\begin{widetext}
\begin{multline}
V^{2b}(2)(m)\equiv\frac{1}{\kappa^2}{}_1\langle m|\mathcal{V}^{2b}_{ij}(2)|m\rangle_1=-\sum_{n=1}^{\infty}\frac{1}{2 n}\biggl[{}_r\langle n,m+n| v_{12} |0,m\rangle_r^2+{}_r\langle n+2,m+n| v_{12} |2,m\rangle_r^2\\
-\frac{1}{2^{n+1}}\Bigl({}_r\langle n,m+n| v_{12} |0,m\rangle_r\sqrt{(n+2) (n+1)/2}-{}_r\langle n+2,m+n| v_{12} |2,m\rangle_r\Bigr)^2\biggr].
\end{multline}
\end{widetext}

For $\mathcal{V}^{2b}(0)$, appearing in Eq.~\eqref{V2bSLL}, we have the equivalent first quantization form

\be\label{fstQV0}
\mathcal{V}^{2b}_{ij}(0)=-\frac{\kappa^2}{2}v_{ij}\biggl[\sum_{1,2}|1_i 2_j\rangle \langle1_i 2_j| \delta_{n_1,0}\delta_{n_2,0}\biggr]v_{ij}.
\ee

\noindent This form leads to the following analytic expression for the Haldane pseudopotentials:

\begin{multline}
V^{2b}(0)(m)\equiv\frac{1}{\kappa^2}{}_1\langle m|\mathcal{V}^{2b}_{ij}(0)|m\rangle_1\\
=-\frac{1}{4}{}_r\langle 2,m| v_{12} |0,m-2\rangle_r^2.
\end{multline} 

We convert the remaining two-body interactions into first quantization using a slightly different technique, along similar lines to the original computation of the conventional Haldane pseudopotentials in the $n=1$ LL~\cite{Haldane1987}. We write the interaction as a Fourier sum

\be
v_{ij}=\sum_q v_q e^{iq\cdot(r_i-r_j)},
\ee 

\noindent where $\sum_q$ abbreviates $1/A\sum_q$. 

For $\mathcal{V}^{2b}_a(1)$ the interaction matrix appearing in Eq.~\eqref{V2bSLL2} can be seen to correspond to the following first quantized interaction:

\begin{multline}
\mathcal{V}^{2b}_{a,ij}(1)=-2 \kappa^2 f\sum_{q_1,q_2} v_{q_1} v_{q_2} e^{iq_1\cdot r_i-iq_2\cdot r_j}\\
\times \biggl[\sum_{5,6}\frac{\tau(n_5,n_6)}{n_6}\langle6|e^{iq_2\cdot r}| 5\rangle \langle 5|e^{-iq_1\cdot r}|6\rangle\biggl],
\end{multline}

\noindent where $|5\rangle$ abbreviates single-particle state $|n_5m_5\rangle$ (similarly for $|6\rangle$), $\tau(n_5,n_6)$ restricts $n_6\geq1$ and $n_5=0$, and $f$ is the spin multiplicity. Then using the properties of the matrix elements of the single-particle density operator~\cite{cond-mat/9410047,Giuliani2005}, one finds

\be\label{1stQVa}
\mathcal{V}^{2b}_{a,ij}(1)=-\frac{\kappa^2 f}{\pi}\int_q  v_q^2 \sum_{n=1}^{\infty}\frac{|\mathcal{F}_{0n}(q)|^2}{n} \ e^{iq\cdot(r_i-r_j)},
\ee

\noindent where $|\mathcal{F}_{0n}(q)|^2=|q|^{2n}e^{-|q|^2/2}/(2^n n!)$ is the modulus squared of the density form factors between the $0$ and $n>0$ Landau levels~\cite{cond-mat/9410047,Giuliani2005} (explicitly presented in Eq.~\eqref{Ffactor} below), and $\int_q=\int d^2 q/(2\pi)^2$. The Haldane pseudopotentials are then found to be

\begin{multline}
V^{2b}_{a}(1)(m)=-\frac{f}{2 \pi^2}\int_0^\infty dq \ q  v^2_q [L_1(q^2/2)]^2 L_m(q^2)\\
\times e^{-3 q^2/2}  \bigl[-\gamma+Ei(q^2/2)-Ln(q^2/2)\bigr],
\end{multline}

\noindent where $\gamma\approx 0.5772$ is the Euler-Mascheroni constant, $Ei(x)=-\int_{-x}^{\infty}dt e^{-t}/t$ is the exponential integral, $L_n$ are the Laguerre polynomials, and $Ln$ is the natural logarithm. We substituted the explicit form of the Coulomb potential, $v_q=2\pi/q$, in this expression to obtain the numbers listed in Table~\ref{tabVm2SLL}.

It is interesting to note that this expression is the $\kappa^2$ term of the well-known RPA screened potential for a completely filled $n=0$ LL, as we demonstrate below. The static screened RPA potential is,

\be
v_q^{RPA}=\kappa v_q+\kappa^2 v_q^2 \chi^0_{q}+\mathcal{O}(\kappa^3),
\ee 

\noindent with $\chi^0_{q}$ the static density-density response function. For a completely filled $n=0$ LL, the static density-density response function is~\cite{Giuliani2005}

\be\label{chi}
\chi^0_q=\frac{f}{2\pi}\sum_{n\neq n'}\frac{f_n-f_n'}{n-n'}|\mathcal{F}_{nn'}(q)|^2=-\frac{f}{2\pi}\sum_{n=1}^{\infty}\frac{|\mathcal{F}_{0n}(q)|^2}{n},
\ee

\noindent where the Fermi occupation factor, $f_n$, is $1$ for $n=0$ and $0$ otherwise. The density form factors $\mathcal{F}_{nn'}(q)$ are given in Eq.~\eqref{Ffactor} below. It is clear from Eq.~\eqref{chi}, that the $\kappa^2$ term of $v_q^{RPA}$ is equivalent to $\mathcal{V}^{2b}_{a,ij}(1)$ appearing in Eq.~\eqref{1stQVa}. 
 
$\mathcal{V}^{2b}_b(1)$, appearing in Eq.~\eqref{V2bSLL2}, can be seen to correspond to the following first quantized interaction,

\begin{multline}\label{fstQV1b}
\mathcal{V}^{2b}_{b,ij}(1)=\sum_{q_1,q_2}v_{q_1}v_{q_2}e^{iq_2\cdot r_i}\biggl[\sum_{6,n_6\geq1}\frac{| 6_i\rangle \langle 6_i|}{n_6}\biggr]e^{iq_1\cdot r_i}\\
\times e^{-iq_1\cdot r_j}\biggl[\sum_{5,n_5=0}| 5_j\rangle \langle 5_j|\biggr]e^{-iq_2\cdot r_j}+\{i\Leftrightarrow j\},
\end{multline}

\noindent where $| 6_i\rangle \langle 6_i|$ operates only on particle $i$, and $| 5_j\rangle \langle 5_j|$ only on $j$. It is not transparent in this representation that $\mathcal{V}^{2b}_{b,ij}(1)$ has translational and rotational invariance. These properties can be made manifest if we write the position operators, $r$, in terms of the mechanical momentum, $\pi$, and guiding center, $c$, coordinates, from the following equations: 

\begin{equation}\label{xx}
\begin{split}
\pi& =p+\frac{e}{c} A(r),\\
c& =r-\hat{z}\times \pi.\\
\end{split}
\end{equation}

\noindent The mechanical momentum and guiding center coordinates are related to the lowering operators as $a=(\pi_{x}+i \pi_{y})/\sqrt{2}$, $b=(c_{x}-i c_{y})/\sqrt{2}$~\cite{cond-mat/9410047,Giuliani2005}. 

The interaction in Eq.~\eqref{fstQV1b}, projected into the $n=1$ LL, can be written as

\begin{multline}
\mathcal{V}^{2b}_{b,ij}(1)=\int_{q_1}\int_{q_2}v_{q_1}v_{q_2}e^{i(q_1-q_2)\cdot (c_i-c_j)}\mathcal{F}_{10}(-q_1)\mathcal{F}_{01}(q_2)\\
\times \biggl[\sum_{n=1}^{\infty}\frac{\mathcal{F}_{1n}(-q_2)\mathcal{F}_{n1}(q_1)}{n}\biggr]+\{i\Leftrightarrow j\},
\end{multline}

\noindent where $\mathcal{F}_{n'n}(q)\equiv \langle n'|\exp(iq\cdot \hat{z}\times\pi)|n\rangle $ are the density form factors, given by~\cite{cond-mat/9410047,Giuliani2005}

\be\label{Ffactor}
\mathcal{F}_{n'n}(q)=\sqrt\frac{n!}{n'!}\biggl(\frac{q_x+iq_y}{\sqrt{2}}\biggr)^{n'-n} e^{-\frac{|q|^2}{4}}L_{n}^{n'-n}\Bigl(\frac{|q|^2}{2}\Bigr),
\ee

\noindent for $n'\geq n$, and $\mathcal{F}_{n'n}(-q)=\mathcal{F}_{nn'}^*(q)$. Combinig this with the analogue relation for the guiding center coordinates,

\be\label{CFfactor}
{}_r\langle0,m|e^{iq\cdot(c_i-c_j)}|0,m\rangle_r=e^{-\frac{|q|^2}{2}} L_m(|q|^2),
\ee

\noindent one finds the Haldane pseudopotentials for $\mathcal{V}^{2b}_{b}$ listed in Table~\ref{tabVm2SLL}. 

For $\mathcal{V}^{2b}_c(1)$, appearing in Eq.~\eqref{V2bSLL2}, following an analogous procedure as the one just outlined for $\mathcal{V}^{2b}_b(1)$, one finds the following first quantized representation for the interaction projected into the $n=1$ LL:

\begin{multline}\label{V2bcapp}
\mathcal{V}^{2b}_{c,ij}(1)=4\int_{q_1}\int_{q_2}v_{q_1}v_{q_2}e^{iq_1\cdot (c_i-c_j)}\\
\times \Re\biggl\{e^{i\hat{z}\cdot q_2\times q_1}\mathcal{F}_{01}(-q_2)\mathcal{F}_{11}(-q_1)\biggl[\sum_{n=1}^{\infty}\frac{\mathcal{F}_{1n}(q_2)\mathcal{F}_{n0}(q_1)}{n}\biggr]\biggr\}.
\end{multline}

\noindent Combining Eq.~\eqref{V2bcapp} with Eq.~\eqref{CFfactor} one obtains the Haldane pseudopotentials listed in Table~\ref{tabVm2SLL} for $\mathcal{V}^{2b}_{c}(1)$.

Finally, the pseudopotentials corresponding to the effective interaction in the spin polarized $n=0$ LL, appearing in Eq.~\eqref{V2bo}, can be obtained similarly to those for $\mathcal{V}^{2b}_a(1)$, and read as,

\begin{multline}
V^{2b}_o(m)=-\frac{f_o}{2 \pi^2}\int_0^\infty dq \ q  v^2_q L_m(q^2)\\
\times e^{-3 q^2/2}  \bigl[-\gamma+Ei(q^2/2)-Ln(q^2/2)\bigr],
\end{multline}

\noindent where $f_o$ is the number of fully occupied flavors ($f_o<f$). The numbers associated with this interaction are listed in Table~\ref{tabVPFLL}.

As an independent consistency check, we have computed the Haldane pseudopotentials for the $n=1$ LL appearing in Table~\ref{tabVm2SLL}, for the leading spin triplet states (odd $m$), in a more direct numerical approach using the two-body interaction matrices that appear in Eqs.~\eqref{V2bSLL} and~\eqref{V2bSLL2}, and verified they coincide with those computed through the semi-analytical formulae discussed in this Appendix. 

We have presented our formulae in a manner that can be easily modified to incorporate cases other than pure Coulomb interactions, with the only requirement of rotational invariance. Only Eq.~\eqref{coulmatrix} makes explicit use of the form of the Coulomb interaction.

%


\end{document}